\documentclass[twocolumn,aps,pre,amsmath,showpacs]{revtex4}
\usepackage{graphicx}

\newcommand{\be}{\begin{equation}}
\newcommand{\ee}{\end{equation}}
\newcommand{\bea}{\begin{eqnarray}}
\newcommand{\eea}{\end{eqnarray}}

\begin{document}

\title{Idealized glass transitions for a system of dumbbell molecules}
\author{S.-H. Chong and W.~G{\"o}tze}
\affiliation{Physik-Department, Technische Universit{\"a}t M{\"u}nchen,
85747 Garching, Germany}
\date{\today, Phys. Rev. E, in press}

\begin{abstract}

The mode-coupling theory for ideal glass transitions in simple
systems is generalized to a theory for the glassy dynamics of 
molecular liquids using the density fluctuations of the sites
of the molecule's constituent atoms as the basic structure
variables.
The theory is applied to calculate the liquid-glass phase 
diagram and the form factors for the arrested structure of
a system of symmetric dumbbells of fused hard spheres.
The static structure factors, which enter the equations of 
motion as input, are calculated as function of the packing 
fraction $\varphi$ and the molecule's elongation $\zeta$ within
the reference-interaction-site-model and Percus-Yevick theories.
The critical packing fraction $\varphi_{c}$ for the glass 
transition is obtained as non-monotone function of $\zeta$
with a maximum near $\zeta = 0.43$.
A transition line is calculated separating a 
small-$\zeta$-glass phase with ergodic dipole motion 
from a large-$\zeta$-glass phase where also the reorientational
motion is arrested.
The Debye-Waller factors at the transition are found to be
somewhat larger for sufficiently elongated systems
than those for the simple hard-sphere system,
but the wave-number dependence of the glass-form factors
is quite similar.
The dipole reorientations for $\zeta \ge 0.6$ are
arrested as strongly as density fluctuations with wave vectors
at the position of the first sharp diffraction peak.

\end{abstract}


\maketitle

\section{Introduction}
\label{sec:1}

The mode-coupling theory (MCT) for idealized liquid-glass transitions
has been proposed originally as a microscopic approximation theory for 
the dynamics of simple liquids~\cite{Bengtzelius84}.
The MCT equations formulate the idea that correlation functions for
density fluctuations have to be evaluated self consistently with 
the correlation functions for force fluctuations.
The derived equations require the static structure factor as input,
which is anticipated to be a smooth function of the wave vector
and of control parameters like the packing fraction $\varphi$.
The equations exhibit a bifurcation singularity for certain
values of the control parameters, say, for $\varphi = \varphi_{c}$.
For $\varphi < \varphi_{c}$, the solutions describe ergodic liquid
dynamics, while for $\varphi \ge \varphi_{c}$ nonergodic dynamics
is obtained describing an amorphous solid.
The arrested glass structure for $\varphi \ge \varphi_{c}$ 
is characterized by glass form factors,
also referred to as nonergodicity parameters.
They generalize the concept of the Edwards-Anderson parameter
from the theory of spin glasses~\cite{Edwards75};
they can be determined in scattering experiments and 
molecular-dynamics-simulation studies.
The MCT equations can be solved by asymptotic expansion
using, e.g., $| \varphi - \varphi_{c} |$ as a small
parameter~\cite{Franosch97,Fuchs98}. 
The leading order results establish universal results for the
glassy dynamics~\cite{Goetze92}. 
Anticipating that these universal formulas are valid also for
mixtures and for 
molecular glass-forming systems, extensive tests of the
MCT with data from experiments and simulations have been
carried out during the past ten years~\cite{Goetze99,Kob99}. 
Due to the invention of improved spectrometers and 
progress in simulation techniques, the work of testing
MCT is still an active field.
Let us mention as particularly impressive recent examples
the studies with the optical Kerr effect~\cite{Torre00,Hinze00b},
the depolarized-light-scattering work for toluene~\cite{Wiedersich00},
the quantitative tests of the form factors for silica~\cite{Sciortino01},
and the scaling-law analysis of simulation data for a polymer
model~\cite{Bennemann99b}.
The indicated tests suggest that MCT deals properly with
some essential features of glass-forming systems.

There is a problem in the tests of the universal MCT formulas:
the range of validity of these leading order asymptotic results
is not universal.
For example, the time interval for the density-fluctuation
decay according to von Schweidler's power law depends non-trivially
on the wave vector of the fluctuations.
Fitting data by a power law for times outside the regime of
validity of the asymptotic law may be possible but can yield
misleading conclusions~\cite{Sciortino99}.
The range of validity of the leading order result can be 
determined by calculating the leading corrections or by 
comparing with the numerical solutions of the full
equations of motion~\cite{Franosch97,Fuchs98}.
But for such discussions one has to analyze the complete
equations of motion, 
i.e. one needs an understanding of the
microscopic details of the system.
Thus, there is the necessity to extend the MCT so that models can 
be analyzed which describe the experimental situation closely.
This is the motivation for the present paper where 
MCT shall be extended to molecular liquids and where this extension
shall be exemplified for a hard-dumbbell system.

Extensions of MCT to molecular systems have been studied already,
generalizing the concept of a density-fluctuation correlator
to the one of infinite matrices of correlation functions formed
with tensor-density 
fluctuations~\cite{Schilling97,Theis98,Fabbian99b,Winkler00,Theis00,Letz00,Theenhaus01,Franosch97c,Goetze00c}.
The results calculated for the glass-form factors for a model
of water~\cite{Fabbian99b,Theis00}
and for a liquid of linear molecules~\cite{Winkler00}
could be used to explain simulation data quantitatively.
Promising results for anomalous oscillation 
spectra for a dipolar-hard-sphere system have been 
calculated~\cite{Theenhaus01}.
For the model of a dilute solute of linear molecules
in a solvent of spherical particles, the MCT equations could be
fully solved~\cite{Franosch97c,Goetze00c}.
The solutions were used to demonstrate the applicability of the
universal formulas also for reorientational motion and to explain
the characteristic difference between the $\alpha$-peaks for
dielectric-loss and depolarized-light-scattering spectra,
as they have been observed in some experiments for 
van-der-Waals liquids.
The cited work shows that MCT studies may contribute to 
the understanding of glassy dynamics which is beyond the
implications of universality.

The MCT equations based on the tensor-density description
of molecular systems 
have a different mathematical structure than the ones
studied so far.
It is unclear whether the bifurcation dynamics of these
equations exhibits the same universal laws as derived
within the MCT for atomic systems.
It is not obvious that codes can be developed for the
numerical solutions of these equations within the
regime of glassy dynamics. 
Therefore it was suggested to base the MCT for molecular systems
on the site-representation~\cite{Chong98b,Chong01,Chong01b}.
This leads to equations with $n$-by-$n$ matrices where
$n$ is the number of atoms producing the force centers in the
molecules.
For the simple case of a dilute solution of linear molecules,
it was shown that this approach yields results~\cite{Chong01}
in semiquantitative agreement with the much more involved
tensor-density theory~\cite{Goetze00c}.
In the present paper, this work shall be continued with the
intention to demonstrate a complete set of results for the
glassy dynamics of a system of linear molecules.

The paper is organized as follows.
The basic general MCT equations are obtained in Appendix~\ref{appen:A}
by modification and generalization of the previous 
work~\cite{Chong98b,Chong01}.
They are specialized in Sec.~\ref{sec:2} to a formulation
of the equations of motion for the coherent and incoherent
density correlation functions for the symmetric-hard-dumbbell system.
The static structure factors, which determine the 
mode-coupling coefficients, are evaluated within the
RISM theory.
To analyze their features in Sec.~\ref{sec:3},
they are decomposed in their various angular momentum
contributions which are evaluated within the Percus-Yevick theory.
Section~\ref{sec:4} explains the phase diagram for the system
and the glass-form factors.
The conclusions (Sec.~\ref{sec:5}) summarize the results while the
discussion of the correlation functions is 
left for a following paper~\cite{Chong-MCT-dumbbell-2}.

\section{A mode-coupling theory for a system of symmetric dumbbells}
\label{sec:2}

\subsection{The model}

A system of $N$ rigid dumbbell molecules distributed with 
density $\rho$ is considered.
The molecule shall be described within the interaction-site 
formalism~\cite{Chandler72,Hansen86}, where the constituent 
atoms shall be called atoms $A$ and $B$. 
Let ${\vec r}_{i}^{\, a}$, $a = A$ or $B$, denote the position vectors
of the atoms in the $i$th molecule, so that 
$L = | {\vec r}_{i}^{\, A} - {\vec r}_{i}^{\, B} |$ denotes
the distance between the two interaction sites. 
Vector ${\vec e}_{i} = ({\vec r}_{i}^{\, A} - {\vec r}_{i}^{\, B}) / L$ 
abbreviates the axis of the $i$th molecule. 
Denoting the mass of atom $a$ as $m_{a}$, the total mass 
$M = m_{A} + m_{B}$ and the moment of inertia $I = m_{A} m_{B} L^2 / M$
determine the thermal velocities 
$v_{T} = \sqrt{k_{B} T / M}$ and 
$v_{R} = \sqrt{k_{B} T / I}$ for the molecule's translation and rotation,
respectively. 
Here $T$ denotes the temperature. 
Let us introduce also the center-of-mass position
${\vec r}_{i}^{\, C} = 
(m_{A} {\vec r}_{i}^{\, A} + m_{B} {\vec r}_{i}^{\, B}) / M$ 
and the coordinates
$z_{a}$ of the atoms along the molecule's axis: 
$z_{A} = L (m_{B} / M)$, 
$z_{B} = - L (m_{A} / M)$. 
The basic structural variables are the
two interaction-site-density fluctuations
for wave vectors ${\vec q}$: 
\begin{equation}
\rho^{a}_{\vec q} = \sum_{i=1}^{N} 
\exp ( i {\vec q} \cdot {\vec r}_{i}^{\, a} ),
\quad a = A \mbox{ or } B.
\label{eq:rho-def}
\end{equation}
The site-site static structure factors 
$S_{q}^{ab} = \langle \rho_{\vec q}^{a \, *} \rho_{\vec q}^{b} \rangle / N$
provide the simplest information on the equilibrium structure of the
system.
Here $\langle \cdots \rangle$ denotes canonical averaging. 
Because of isotropy, $S_{q}^{ab}$ depends only
on the wave number $q = | \, {\vec q} \, |$.
The site-site static structure factor $S_{q}^{ab}$ consists of the 
intramolecular and the intermolecular parts.
The former is denoted as $w_{q}^{ab}$; 
for a rigid dumbbell molecule it is given by
$w^{ab}_{q} = \delta^{ab} + (1 - \delta^{ab}) \, j_{0}(qL)$.
Here and in the following $j_\ell (x)$ denotes the spherical
Bessel function of index $\ell$.
The static structure factors $S_{q}^{ab}$
shall be combined to a two-by-two matrix ${\bf S}_{q}$,
and similar matrix notation will be used 
for other site-site correlation functions. 
The site-site Ornstein-Zernike equation~\cite{Chandler72,Hansen86},
${\bf S}_{q} = [{\bf w}_{q}^{-1} - \rho {\bf c}_{q}]^{-1}$,
relates $S_{q}^{ab}$ to the site-site direct correlation function
$c_{q}^{ab}$.

The structural dynamics of the system shall be described by the
interaction-site-density correlators 
\begin{equation}
F^{ab}_{q}(t) = 
\langle \rho^{a}_{\vec q}(t)^{*} \rho^{b}_{\vec q}(0) \rangle / N.
\label{eq:Fab-def}
\end{equation}
These are real even functions of time obeying 
$F_{q}^{ab} (t) = F_{q}^{ba} (t)$.
The short-time expansion can be written as
\begin{equation}
{\bf F}_{q}(t) = {\bf S}_{q} - 
{\textstyle \frac{1}{2}} \, q^{2} \, {\bf J}_{q} \, t^{2} +
{\bf O}(t^{4}).
\label{eq:Fab-short-time}
\end{equation}
The continuity equation reads 
$\dot{\rho}_{\vec q}^{a} = i {\vec q} \cdot {\vec j}_{\vec q}^{a}$, 
where the longitudinal current fluctuation is given by 
$\vec j_{\vec q}^a = \sum_{i} {\vec v}_{i}^{\, a} 
\exp(i {\vec q} \cdot {\vec r}_{i}^{\, a})$
with ${\vec v}_{i}^{\, a}$ denoting the
velocity of atom $a$ in the $i$th molecule. 
Therefore, one gets 
$J_q^{ab} = \langle 
({\vec q} \cdot {\vec j}_{\vec q}^{\, a})^* 
({\vec q} \cdot {\vec j}_{\vec q}^{\, b}) 
\rangle / N q^2$, whose explicit expressions for a rigid
dumbbell molecule are given by~\cite{Chong98}
\begin{equation}
J^{ab}_{q} = 
v_{T}^{2} \, w^{ab}_{q} +
v_{R}^{2} \, 
( {\textstyle \frac{2}{3}} z_{a} z_{b} ) \,
[ \delta^{ab} + (1-\delta^{ab}) \, (j_{0}(qL) + j_{2}(qL)) ].
\label{eq:Jab}
\end{equation}

The dynamics of the tagged molecule
shall also be considered.
It is described by the self part of the 
interaction-site-density correlators
\be
F_{q,s}^{ab}(t) = 
\langle \rho^{a}_{{\vec q},s}(t)^{*} \rho^{b}_{{\vec q},s}(0) \rangle.
\label{eq:Fs-def}
\end{equation}
Here $\rho_{{\vec q},s}^{a} = \exp( i {\vec q} \cdot {\vec r}_{s}^{\, a})$
with ${\vec r}_{s}^{\, a}$ denoting the position vector of atom $a$
in the tagged molecule. 
The short-time expansion of the correlator
${\bf F}_{q,s}(t)$ is given by Eq.~(\ref{eq:Fab-short-time}) with 
${\bf S}_{q}$ replaced by ${\bf w}_{q}$.
The same function $J_{q}^{ab}$ determines the short-time dynamics
of $F_{q,s}^{ab}(t)$ since the velocities of different molecules
at the same time are statistically independent.

For later convenience, it shall be shown here 
how the correlation functions in the
interaction-site representation can be expressed in terms of the
ones in the tensor-density description.
Following the convention in Refs.~\onlinecite{Franosch97c} and
\onlinecite{Goetze00c}, 
coherent tensor-density fluctuations 
$\rho_{\ell}^{m}({\vec q} \,)$ for the angular-momentum index $\ell$
and the helicity index $m$ shall be defined by
decomposing the $i$th molecule's position variable in plane waves 
$\exp( i {\vec q} \cdot {\vec r}_{i}^{\, C})$ 
for the center of mass ${\vec r}_{i}^{\, C}$ 
and in spherical harmonics $Y_{\ell}^{m}({\vec e}_{i})$ for the 
orientation vector ${\vec e}_{i}$:
\begin{equation}
\rho_{\ell}^{m}({\vec q} \,) = i^{\ell} \, \sqrt{4 \pi} 
\sum_{i=1}^{N}
\exp ( i {\vec q} \cdot {\vec r}_{i}^{\, C}) \,
Y_{\ell}^{m}({\vec e}_{i}).
\label{eq:rho-tensor-def}
\end{equation}
The structural dynamics is described by the matrix of correlators
\be
\Phi_{\ell \ell'}^{m}(q,t) = 
\langle \rho_{\ell}^{m}({\vec q}_{0},t)^{*} 
\rho_{\ell'}^{m}({\vec q}_{0},0) \rangle / N; \quad
{\vec q}_{0} = (0,0,q). 
\ee
The general correlators
$\langle \rho_{\ell}^{m}({\vec q},t)^{*} \rho_{\ell'}^{m'}({\vec q},0) \rangle$
can be written as linear combination of the functions
$\Phi_{\ell \ell'}^{m}(q,t)$;
they vanish for $m \ne m'$ if ${\vec q} = {\vec q}_{0}$~\cite{Franosch97c}.
In particular, 
the equilibrium structure is described by the static
correlation functions
\begin{equation}
S_{\ell \ell'}^{m}(q) = 
\langle \rho_{\ell}^{m}({\vec q}_{0})^{*} 
\rho_{\ell'}^{m}({\vec q}_{0}) \rangle / N.
\label{eq:Sq-tensor}
\end{equation}
Since the position vectors of the interaction sites can be written as 
${\vec r}_{i}^{\, a} = {\vec r}_{i}^{\, C} + z_a {\vec e}_{i}$, 
the Rayleigh expansion of the exponential in Eq.~(\ref{eq:rho-def}) 
yields the formula
\begin{equation}
\rho^{a}_{{\vec q}_{0}} = \sum_{\ell} \sqrt{2 \ell + 1} \, 
j_{\ell}(q z_{a}) \, \rho_{\ell}^{0}({\vec q}_{0}).
\label{eq:rho-site-tensor}
\end{equation}
Substitution of this expression into Eq.~(\ref{eq:Fab-def})
leads to an expression for the density correlators
in the site representation in terms of those in the tensor-density
description:
\be
F_{q}^{ab}(t) = \sum_{\ell, \ell'}
\sqrt{(2\ell+1)(2\ell'+1)} \,
j_{\ell}(q z_{a}) \, j_{\ell'}(q z_{b}) \,
\Phi_{\ell \ell'}^{0}(q,t).
\label{eq:Fq-decom}
\ee
In particular, the site-site static structure factors
$S_{q}^{ab}$ are related to the tensorial ones via 
\be
S_{q}^{ab} = \sum_{\ell, \ell'}
\sqrt{(2\ell+1)(2\ell'+1)} \,
j_{\ell}(q z_{a}) \, j_{\ell'}(q z_{b}) \,
S_{\ell \ell'}^{0}(q).
\label{eq:Sq-decom}
\ee
Similarly, one obtains formulas relating tagged-molecule
correlators in the site representation and those in the tensor-density
description:
\bea
F_{q,s}^{ab}(t) &=&
\sum_{\ell, \ell'}
\sqrt{(2\ell+1)(2\ell'+1)} 
\nonumber \\
& &
\qquad
\times \,
j_{\ell}(q z_{a}) \, j_{\ell'}(q z_{b}) \,
\Phi_{s,\ell \ell'}^{0}(q,t),
\label{eq:Fqs-decom}
\eea
where $\Phi_{s,\ell \ell'}^{0}(q,t)$ denotes the self part of
$\Phi_{\ell \ell'}^{0}(q,t)$.
Since $\Phi_{s,\ell \ell'}^{0}(q,0) = \delta_{\ell \ell'}$,
one gets
\be
w_{q}^{ab} = \sum_{\ell} (2\ell + 1) \,
j_{\ell}(q z_{a}) \, j_{\ell}(q z_{b}).
\label{eq:wq-decom}
\ee

In the present paper, a system of symmetric dumbbell molecules,
consisting of two fused hard spheres of diameters 
$d_{A} = d_{B} = d$ and masses $m_{A} = m_{B} = M/2$,
shall be considered.
The elongation parameter $\zeta = L/d$ quantifies the bond length.
All equilibrium properties of such a hard-dumbbell system with
a fixed elongation are specified by the packing fraction
\be
\varphi = \rho V_{0}, \quad
V_{0} = \frac{\pi}{6} d^{3} 
\biggl( 1 + \frac{3}{2} \zeta - \frac{1}{2} \zeta^{3} \biggr).
\ee
Here $V_{0}$ is the volume of a dumbbell molecule.
Throughout the rest of this paper, the diameter of the constituent
atom is chosen as the unit of length, $d = 1$. 

For a symmetric system, 
there are only two independent density correlators,
since $F^{AA}_{q}(t) = F^{BB}_{q}(t)$.
It is convenient to perform an orthogonal transformation to fluctuations
of total number density
$\rho_{\vec q}^{N}$ and 
``charge'' density
$\rho_{\vec q}^{Z}$: 
\begin{subequations}
\label{eq:rho-NZ}
\begin{equation}
\rho_{\vec q}^{x} =
( \rho^{A}_{\vec q} \pm \rho^{B}_{\vec q} ) \, / \, \sqrt{2}, \quad
x = N \mbox{ or } Z.
\label{eq:rho-NZ-a}
\end{equation}
The transformation matrix ${\bf P} = {\bf P}^{-1}$ reads
\begin{equation}
{\bf P} = \frac{1}{\sqrt{2}}
\left(
\begin{array}{rr}
1 &  1 \\
1 & -1 
\end{array}
\right).
\label{eq:rho-NZ-b}
\end{equation}
It diagonalizes the matrices ${\bf S}_{q}$, 
${\bf w}_{q}$, and ${\bf J}_{q}$: 
\bea
& &
( {\bf P} \, {\bf S}_{q} \, {\bf P} )^{xy} =
\delta^{xy} \, S_{q}^{x}, \quad
S_{q}^{x} = S_{q}^{AA} \pm S_{q}^{AB},
\label{eq:rho-NZ-c}
\\
& &
( {\bf P} \, {\bf w}_{q} \, {\bf P} )^{xy} =
\delta^{xy} \, w_{q}^{x}, \quad
w_{q}^{x} = 1 \pm j_{0}(q \zeta),
\label{eq:rho-NZ-d}
\\
& &
( {\bf P} \, {\bf J}_{q} \, {\bf P} )^{xy} =
\delta^{xy} \, 
\{ 
v_{T}^{2} \, w_{q}^{x} 
\nonumber \\
& &
\qquad \qquad  
+ \,
{\textstyle \frac{1}{6}} \, v_{R}^{2} \, \zeta^{2} \,
[1 \mp (j_{0}(q \zeta) + j_{2}(q \zeta))] \},
\label{eq:rho-NZ-e}
\eea
\end{subequations}
where $x,y = N$ or $Z$.
Also the matrix of density correlators is diagonalized.
Introducing the density correlators $\phi_{q}^{x}(t)$
normalized to $\phi_{q}^{x}(t=0)=1$, one gets 
\begin{subequations}
\label{eq:phi-NZ}
\bea
& &
\phi^{x}_{q}(t) =
\langle \rho_{\vec q}^{x}(t)^{*} \rho_{\vec q}^{x}(0) \rangle \, / \, 
N S_{q}^{x},
\\ 
& &
( {\bf P} \, {\bf F}_{q}(t) \, {\bf P} )^{xy} =
\delta^{xy} \, \phi^{x}_{q}(t) \, S_{q}^{x},
\eea
\end{subequations}
and similar equations hold for the normalized
tagged-molecule correlators $\phi_{q,s}^{x}(t)$: 
\begin{subequations}
\label{eq:phi-sNZ}
\bea
& &
\phi^{x}_{q,s}(t) = 
\langle 
\rho_{{\vec q},s}^{x}(t)^{*} 
\rho_{{\vec q},s}^{x}(0) \rangle \, / \, w_{q}^{x}, 
\\
& &
( {\bf P} \, {\bf F}_{q,s}(t) \, {\bf P} )^{xy} =
\delta^{xy} \, \phi^{x}_{q,s}(t) \, w_{q}^{x}.
\eea
\end{subequations}

There is an additional property due to the symmetry of the molecule.
Since the intermolecular parts of $S_{q}^{AA}$ 
and $S_{q}^{AB}$ are the same, one gets
$S_{q}^{Z} = w_{q}^{Z}$.
A similar reasoning for the charge-density correlators leads to
\be 
\phi_{q}^{Z}(t) = \phi_{q,s}^{Z}(t).
\label{eq:phi-Z-symm}
\ee 
For a system of symmetric dumbbell molecules,
there is only one independent coherent density correlator,
viz. $\phi_{q}^{N}(t)$. 

\subsection{MCT equations for the density correlators}

The MCT equations of motion for the density correlators
consist of an exact Zwanzig-Mori equation and the approximate 
expression for the relaxation kernel in terms of the mode-coupling functional,
whose derivation is described in Appendix~\ref{appen:A}.
For a system of symmetric dumbbells, these equations can be 
simplified considerably~\cite{Chong01}.
Multiplying Eqs.~(\ref{eq:GLE-v-a})--(\ref{eq:MCT-v-c}) from left and right with 
$\bf P$ given by Eq.~(\ref{eq:rho-NZ-b}) and inserting ${\bf 1} = {\bf P P}$ 
between every pair of matrices, all equations are
transformed to diagonal ones. 
Thus, there are two sets of equations, 
one for $\phi_{q}^{N}(t)$
and another for $\phi_{q}^{Z}(t)$.
As explained in connection with Eq.~(\ref{eq:phi-Z-symm}), 
the charge-density correlator $\phi_{q}^{Z}(t)$
is identical to its self part, $\phi_{q,s}^{Z}(t)$, which shall be
treated separately below.
Thus, the only correlator describing the coherent density
fluctuations is the total density correlator $\phi_{q}^{N}(t)$,
whose Zwanzig-Mori equation reads
\begin{subequations}
\label{eq:GLE-N}
\bea
& &
\partial_{t}^{2} \phi^{N}_{q}(t) + (\Omega_{q}^{N})^{2} \, \phi^{N}_{q}(t) 
\nonumber \\
& &
\quad 
+ \,
(\Omega_{q}^{N})^{2} 
\int_{0}^{t} dt' \, m^{N}_{q}(t-t') \, \partial_{t'} \phi^{N}_{q}(t') = 0.
\label{eq:GLE-N-a}
\eea
The characteristic frequency $\Omega_{q}^{N}$, which specifies the
initial decay of the correlator by 
$\phi_{q}^{N} (t) = 1 - \frac{1}{2} (\Omega_{q}^N t)^2 + O (t^4)$, is given by
\bea
& &
(\Omega_{q}^{N})^{2} = q^{2} \, 
\{ v_{T}^{2} \, [1 + j_{0}(q \zeta)] 
\nonumber \\
& &
\qquad
+ \,
   {\textstyle \frac{1}{6}} v_{R}^{2} \zeta^{2} \, [1 - j_{0}(q \zeta) - j_{2}(q \zeta)]
\} \, / \, S_{q}^{N}.
\label{eq:GLE-N-b}
\eea
\end{subequations}
The relaxation kernel reads 
$m_{q}^{N}(t) = 
{\cal F}_{q}^{N} [\phi^{N}(t)]$, 
where Eqs.~(\ref{eq:MCT-v-a})--(\ref{eq:MCT-v-c})
lead to 
\begin{subequations}
\label{eq:MCT-N}
\bea
& &
{\cal F}^{N}_{q}[\tilde{f}] = 
\frac{1}{2} \int d{\vec k} \,
V^{N}({\vec q}; {\vec k}, {\vec p} \,) \, 
\tilde{f}_{k} \, \tilde{f}_{p},
\label{eq:MCT-N-a}
\\
& &
V^{N}({\vec q}; {\vec k}, {\vec p} \,) =
\frac{\rho}{16 \pi^{3} q^{4}}
S_{q}^{N} S_{k}^{N} S_{p}^{N}
\nonumber \\
& &
\qquad \qquad \qquad \qquad
\times \,
\{ {\vec q} \cdot [ {\vec k} c_{k}^{N} + {\vec p} c_{p}^{N} ] \}^{2},
\label{eq:MCT-N-b}
\eea
\end{subequations}
with ${\vec p} = {\vec q} - {\vec k}$ and 
$c_{q}^{N} = 2 c_{q}^{AA}$. 
One gets from Eq.~(\ref{eq:DW-v}) for the nonergodicity parameters
$f_{q}^{N} = \phi_{q}^{N}(t \to \infty)$:
\begin{equation}
f^{N}_{q} = {\cal F}^{N}_{q}[f^{N}] \, / \, 
\{ 1 + {\cal F}^{N}_{q}[f^{N}] \}.
\label{eq:DW-N}
\end{equation}
Notice that Eqs.~(\ref{eq:GLE-N-a}), (\ref{eq:MCT-N}), and (\ref{eq:DW-N}) 
are formally identical to the corresponding equations 
for simple systems~\cite{Franosch97}: 
the difference is in the definition
of the correlators and of the direct correlation functions. 
In particular, one can show that the preceding equations 
(\ref{eq:GLE-N})--(\ref{eq:DW-N}) 
reduce to the ones for simple systems
in both the $\zeta \to 0$ and $\zeta \to \infty$ limits. 

The MCT model for the hard-dumbbell system (HDS) will be defined by
two further technical assumptions.
First, the site-site structure factors $S_{q}^{ab}$ and the direct
correlation functions $c_{q}^{ab}$ are evaluated within 
the reference-interaction-site-model (RISM) integral-equation 
theory~\cite{Chandler72,Hansen86,Lowden73}. 
Second, the wave numbers are discretized to 100 equally spaced values 
$q = 0.2$, $0.6$, $1.0$, $\cdots$, $39.8$.
The details of the transformation of the mode-coupling functional
to a polynomial in the discretized variables can be found in 
Ref.~\onlinecite{Franosch97}.

\begin{figure}
\includegraphics[width=0.9\linewidth]{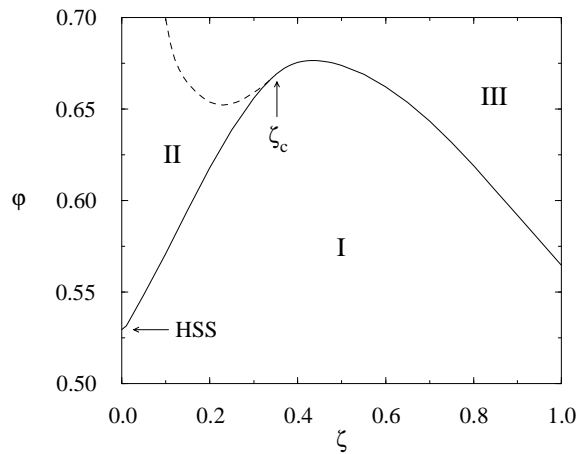}
\caption{
Phase diagram of the symmetric-hard-dumbbell system 
where the packing fraction is denoted by $\varphi$
and the elongation parameter by $\zeta$. 
The solid curve marks the type-$B$ liquid-glass-transition line,
$\varphi_{c} = \varphi_{c}(\zeta)$. 
The dashed curve denotes the type-$A$-transition line
between phases II and III, 
$\varphi_{A} = \varphi_{A}(\zeta)$.
The type-$A$-transition line terminates at the critical
elongation $\zeta_{c} = 0.345$ marked by an arrow.
The horizontal arrow marks the transition point of the
hard-sphere system (HSS).}
\label{fig:phase}
\end{figure}

The discussion of Eq.~(\ref{eq:DW-N}) can follow that considered
previously for simple systems~\cite{Franosch97}. 
For a given $\zeta$, one finds a critical packing fraction
$\varphi_{c} = \varphi_{c}(\zeta)$ so that
$f_{q}^{N} = 0$ for $\varphi < \varphi_{c}$ and
$f_{q}^{N} > 0$ for $\varphi \ge \varphi_{c}$.
Figure~\ref{fig:phase} exhibits the control-parameter plane for our system;
the full line represents the $\varphi_{c}$--versus--$\zeta$ curve.
The regime I, i.e., the states with $(\zeta,\varphi)$ below the
full line, are the liquid states.
For states on and above the line, the density-fluctuation 
dynamics is nonergodic.
It is the purpose of this paper to explain the origin of this
liquid-glass-transition curve and to quantify the arrested glass 
structure. 

Comments on some technical details of our calculations
might be in order.
We solved a set of equations in the RISM theory to obtain $S_{q}^{ab}$
using the non-equally spaced wave-number grids
introduced in Ref.~\onlinecite{Rossky80}.
The resulting $S_{q}^{ab}$ has been subsequently transformed to the 
one on the above-mentioned equispaced grids 
using a cubic spline interpolation~\cite{Press93}.
Our results for the HDS are based on the $S_{q}^{ab}$ so obtained. 
Occasionally, 
we will refer to results for the hard-sphere system (HSS),
i.e. the HDS with $\zeta = 0$. 
For consistency, calculations for the HSS 
have also been done using a static structure factor which is 
based on the same numerical method just mentioned.
However, the numerically obtained structure factor for the HSS is,
due to the interpolation procedure, slightly different from the one
of the analytic Percus-Yevick theory~\cite{Hansen86}. 
Since the transition is sensitively dependent on the structure-factor
peak, this leads to slightly different results for the HSS
from the previous ones reported in Ref.~\onlinecite{Franosch97},
where the analytic Percus-Yevick theory is used.
Typically, the differences are less than 1\%.
Therefore, the results for the HSS based on the two different
static inputs can be regarded as essentially the same.
The nonstandard wave-number grids from Ref.~\onlinecite{Rossky80} 
and the subsequent interpolation procedure
have been adopted because of the following reason.
The present model shall be extended to one where the constituent
atoms carry opposite electrical charges.
Thereby it will be possible to study the interplay of 
steric-hindrance effects and Coulomb-interaction effects.
The method developed in Ref.~\onlinecite{Rossky80} is well suited for 
treating such a system 
in which both the short- and long-ranged interactions are 
simultaneously present.
To have the results for the HDS as reference model, 
it seems adequate to carry out the calculation of the static
input function strictly within the same frame. 

\subsection{MCT equations for the tagged-molecule correlators}

One gets the Zwanzig-Mori equation for the normalized 
tagged-molecule correlator $\phi_{q,s}^{x}(t)$ ($x = N$ or $Z$)
for a symmetric-hard-dumbbell system 
by transforming Eq.~(\ref{eq:GLE-u}) as explained above
for deriving Eq.~(\ref{eq:GLE-N-a}):
\begin{subequations}
\label{eq:GLE-sNZ}
\bea
& &
\partial_{t}^{2} \phi^{x}_{q,s}(t) + 
(\Omega^{x}_{q,s})^{2} \, \phi^{x}_{q,s}(t) 
\nonumber \\
& &
\,\,\,
+ 
(\Omega^{x}_{q,s})^{2} 
\int_{0}^{t} dt' \, m^{x}_{q,s}(t-t') \, 
\partial_{t'} \phi^{x}_{q,s}(t') = 0. 
\label{eq:GLE-sNZ-a}
\eea
The characteristic frequency $\Omega_{q,s}^{x}$ specifies the
initial decay of the correlator by 
$\phi_{q,s}^{x}(t) = 1 - \frac{1}{2} (\Omega_{q,s}^{x} t)^2 + O (t^4)$,
and it is given by
\be
(\Omega_{q,s}^{x})^{2} = q^{2} \,
\{ v_{T}^{2} + {\textstyle \frac{1}{6}} v_{R}^{2} \zeta^{2} \, 
               [1 \mp j_{0}(q \zeta) \mp j_{2}(q \zeta)] \, / \, 
               [1 \pm j_{0}(q \zeta)]                         \}.
\label{eq:GLE-sNZ-b}
\ee
\end{subequations}
The relaxation kernel can be written as 
$m_{q,s}^{x}(t) = 
{\cal F}_{q}^{x} [\phi_{s}^{x}(t), \phi^{N}(t)]$, where 
Eqs.~(\ref{eq:MCT-u-a}) and (\ref{eq:MCT-u-b}) lead to
\be
{\cal F}^{x}_{q,s}[\tilde{f}_{s}^{x}, \tilde{f}] =
\frac{\rho}{16 \pi^{3}} 
\frac{w_{q}^{x}}{q^{2}} 
\int d{\vec k} 
\biggl( \frac{ {\vec q} \cdot {\vec p} }{q} \biggr)^{2} 
(c_{p}^{N})^{2} w_{k}^{x} S_{p}^{N} 
\tilde{f}^{x}_{k,s} {\tilde f}_{p},
\label{eq:MCT-sNZ}
\ee
with ${\vec p} = {\vec q} - {\vec k}$. 
From the long-time limits of 
Eqs.~(\ref{eq:GLE-u}) and (\ref{eq:MCT-u-a}), 
one gets for the nonergodicity parameters
$f_{q,s}^{x} = \phi_{q,s}^{x} (t \to \infty)$: 
\begin{equation}
f^{x}_{q,s} = {\cal F}^{x}_{q,s}[f_{s}^{x},f^{N}] \, / \, 
\{ 1 + {\cal F}^{x}_{q,s}[f_{s}^{x},f^{N}] \}.
\label{eq:DW-sNZ}
\end{equation}
The mathematical structure of the two sets of 
the equations (\ref{eq:GLE-sNZ})--(\ref{eq:DW-sNZ}) for 
$x = N$ and $Z$ is the same as the one studied previously 
for the tagged-particle-density correlator in a simple 
liquid~\cite{Fuchs98}. 
In particular, the set of equations for $x = N$ reduce to the 
one for the tagged-particle-density correlator 
in both the $\zeta \to 0$ and $\zeta \to \infty$ limits. 
In the same limits, the correlator $\phi_{q,s}^{Z}(t)$ becomes
identically zero. 

Equations~(\ref{eq:GLE-sNZ})--(\ref{eq:DW-sNZ}) 
for the tagged symmetric dumbbell
immersed in a liquid of symmetric dumbbells are also 
formally identical to the ones treated in 
Refs.~\onlinecite{Chong01} and \onlinecite{Chong01b} 
for the symmetric dumbbell molecule dissolved in a simple liquid.
This is because the coherent density fluctuations of the surroundings
in the former case is characterized only by the correlator $\phi_{q}^{N}(t)$, 
i.e., a scalar correlator, and this feature is shared
with the latter case.
In analogy to the findings in the previous studies,
one finds a line of transition points $\varphi_{A} = \varphi_{A}(\zeta)$,
provided $\zeta \le \zeta_{c} = 0.345$.
This line, which is shown in dashed in 
Fig.~\ref{fig:phase}, separates glass states 
in regime II and regime III.
In regime II, the reorientational motion is ergodic, i.e., the states
deal with the amorphous analogue of a plastic crystal.
In regime III, also the reorientational motion is nonergodic,
since $f_{q,s}^{Z} > 0$.
Crossing the dashed line by, e.g., increasing $\varphi$, 
$f_{q,s}^{Z}$ change continuously (type-$A$ transition).
Crossing the heavy line, $f_{q}^{N}$ changes discontinuously
(type-$B$ transition).
The interest of the present studies concerns the transition
from the liquid to a glass with all density correlators
arrested, as obtained for $\zeta > \zeta_{c}$ by increasing
$\varphi$.
As a representative situation with strong steric hindrance
for reorientational motion, molecules with $\zeta = 1.0$ shall
be analyzed in detail.
For $\zeta$ approaching $\zeta_{c}$ from above, the steric hindrance
for reorientations weakens and molecules with $\zeta = 0.4$
shall be used to demonstrate this case.

\section{Structure of the relaxed system}
\label{sec:3}

The static structure factor for the total density fluctuations, $S_{q}^{N}$,
is the basic input of our theory.
It quantifies the simplest information on the averaged particle distribution,
anticipating the system to be relaxed in a canonical equilibrium state.
The latter is assumed to be an amorphous one. 
It may be metastable, e.g., with respect to crystallization.
The variation of $S_{q}^{N}$ with changes of the packing fraction
$\varphi$ and the molecule's elongation $\zeta$ provides the
key for explaining the phase diagram in Fig.~\ref{fig:phase}.
Extending earlier work~\cite{Streett76} to the high density regime,
$S_{q}^{N}$ shall be analyzed in this section.

\subsection{Static structure factors and angular correlations}
\label{subsec:Sq}

\begin{figure}
\includegraphics[width=0.7\linewidth]{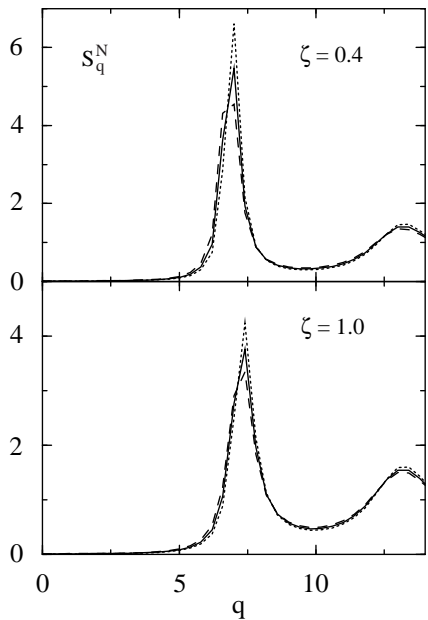}
\caption{
Static structure factor $S_{q}^{N}$ for the total density fluctuations
as function of wave number $q$ 
for the elongations $\zeta = 0.4$ (upper panel)
and $1.0$ (lower panel).
The results refer to packing fractions 
$\varphi = \varphi_{c} (1 + \epsilon)$
with $\epsilon = 0$ (solid lines), 
$\epsilon = - 10^{-5/3}$ (dashed lines), and
$\epsilon = + 10^{-5/3}$ (dotted lines).
Here $\varphi_{c}$ denotes the critical packing fraction;
it is given by $\varphi_{c} = 0.675$ and $0.565$ for
$\zeta = 0.4$ and $1.0$, respectively. 
The first sharp diffraction peak in $S_{q}^{N}$ for $\zeta = 0.4$ 
occurs at $q = 7.0$ in the discretized wave-number grids, 
and its heights are 
4.54, 5.47, and 6.61 with increasing $\varphi$.
The corresponding peak for $\zeta = 1.0$ occurs at $q = 7.4$,
and its heights are 3.33, 3.75, and 4.24 with increasing $\varphi$.
Here and in the following figures the diameter of the
spheres is used as unit of length, $d = 1$.}
\label{fig:Sq-near-phic}
\end{figure}

Figure~\ref{fig:Sq-near-phic} exhibits results for $S_{q}^{N}$  
calculated from the RISM 
theory~\cite{Chandler72,Hansen86,Lowden73}
for the two representative elongations $\zeta = 0.4$ and $1.0$ 
at and near the critical packing fraction $\varphi_{c}(\zeta)$.
For small $q$, the structure factor is small.
Because of the dense packing, the compressibility 
$\kappa_{q} \propto S_{q}^{N}$ for long-wavelength fluctuations
is strongly suppressed.
These fluctuations are irrelevant for the glassy arrest in our system.
The phase diagram does not change more than 1\%
if fluctuations with, say, $q \le 3$ are cut off.
Fluctuations with, say, $q \ge 10$ are relevant, since
$S_{q}^{N}-1$ is of order unity.
But in this regime, the structure factor is not very sensitive with
respect to changes of the density.
Therefore, the liquid-glass transition is driven mainly by the
changes of $S_{q}^{N}$ for $q \approx 7$, i.e. by the
fluctuations with wave vectors near the position of the
first sharp diffraction peak. 
This feature is analogous to the one found in the hard-sphere
system~\cite{Bengtzelius84}.
However, the results shown in Fig.~\ref{fig:Sq-near-phic} 
are for molecular systems in which angular correlations
should play an important role as well. 
This subsection is devoted to discuss how angular correlations
manifest themselves in $S_{q}^{N}$.

To proceed, let us decompose the
$S_{q}^{N}$ in terms of the spherical-harmonic
expansion coefficients $S_{\ell \ell'}^{m}(q)$ defined in 
Eq.~(\ref{eq:Sq-tensor}):
the coefficient $S_{00}^{0}(q)$ describes the static center-of-mass
density fluctuations, 
and the higher coefficients probe the angular correlations.
This decomposition can be derived from Eq.~(\ref{eq:Sq-decom})
by noticing the definition
$S_{q}^{N} = S_{q}^{AA} + S_{q}^{AB}$:
\bea
S_{q}^{N} &=& 
\sum_{\ell,\ell' \rm{: even}}
2 \sqrt{(2\ell+1)(2\ell'+1)} 
\nonumber \\
& &
\qquad \qquad
\times \,
j_{\ell}(q\zeta/2) \, j_{\ell'}(q\zeta/2) 
S_{\ell \ell'}^{0}(q).
\label{eq:Sq-N-decom}
\eea
Here the angular-momentum indices $\ell$ and $\ell'$ take only even numbers
due to the top-down symmetry of the dumbbell molecule.
It is clear 
that the coefficients $S_{\ell \ell'}^{0}(q)$ contain more information
than $S_{q}^{N}$ since the latter
can be expressed in terms of the former, but not {\rm vice versa}.

\begin{figure}
\includegraphics[width=0.7\linewidth]{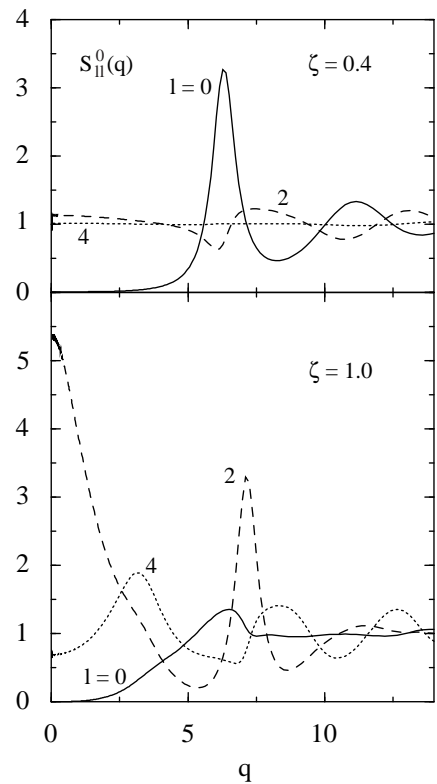}
\caption{
Spherical-harmonic expansion coefficients $S_{\ell \ell}^{0}(q)$ 
of the structure factor 
for the elongations $\zeta = 0.4$ (upper panel) and 
$1.0$ (lower panel) at the
critical packing fraction $\varphi = \varphi_{c}$
for the angular-momentum indices $\ell = 0$, 2, and 4.}
\label{fig:Sq-llm}
\end{figure}

The expansion coefficients $S_{\ell \ell'}^{m}(q)$ 
have been calculated within the Percus-Yevick (PY) 
theory~\cite{Chen71,Gray84} up to the 
angular-momentum-index cutoff $\ell_{\rm cut} = 6$.
For the symmetric dumbbell, this results 
in 30 independent coefficients to be dealt with 
in solving the PY equation. 
The representative results at the critical packing fraction 
for the diagonal coefficients $S_{\ell \ell}^{0}(q)$ are 
shown in Fig.~\ref{fig:Sq-llm}. 
For the small elongation $\zeta = 0.4$,
the density fluctuations for $q \approx 7$
are dominated by those of the center-of-mass
degrees of freedom, $\ell = 0$, 
while contributions from the reorientational 
correlations are rather small.
On the other hand, for the large elongation $\zeta = 1.0$,
the static structure for $q \approx 7$ is primarily caused by
the reorientational function $S_{22}^{0}(q)$, while the center-of-mass
component $S_{00}^{0}(q)$ only shows a weak structure. 
A strong peak at $q \approx 0$ is also seen in the coefficient 
$S_{22}^{0}(q)$ for $\zeta = 1.0$, 
which is a precursor of a nematic instability.
The increased importance of the higher coefficients 
for larger elongations is demonstrated even more clearly
by comparing 
$S_{44}^{0}(q)$ for the two elongations.
These features of the coefficients $S_{\ell \ell}^{0}(q)$
for small and large elongations are in accord with those found in 
Ref.~\onlinecite{Letz99}, 
albeit for fluids of hard ellipsoids
in which the aspect ratio plays a role similar to $(1+\zeta)$.

\begin{figure}
\includegraphics[width=0.7\linewidth]{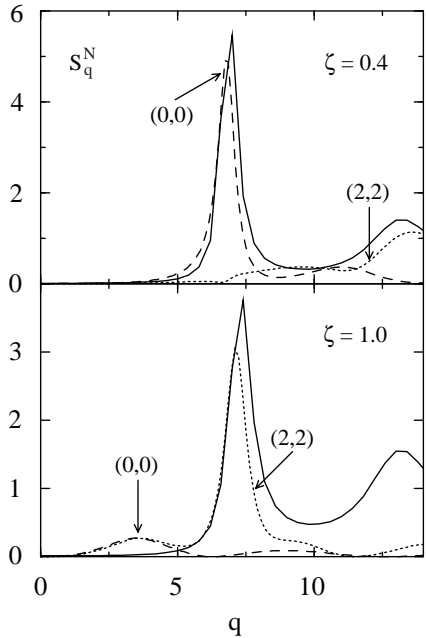}
\caption{
The solid lines denote the static structure factor $S_{q}^{N}$ 
for the total density
fluctuations for the elongation $\zeta = 0.4$ (upper panel)
and $\zeta = 1.0$ (lower panel)
at the critical packing fraction $\varphi = \varphi_{c}(\zeta)$.
The dashed and dotted lines are the results of
decomposition based on 
Eq.~(\protect\ref{eq:Sq-N-decom}),
where the numbers in parentheses $(\ell, \ell)$ indicate the
component in the decomposition (see text).} 
\label{fig:Sq-decom}
\end{figure}

Figure~\ref{fig:Sq-decom} exhibits the decomposition of $S_{q}^{N}$ 
at the critical packing fraction 
based on Eq.~(\ref{eq:Sq-N-decom}). 
The solid lines denote $S_{q}^{N}$ calculated from the RISM
theory, and 
the dashed and dotted lines 
denote the terms in the decomposition using the coefficients
$S_{\ell \ell}^{0}(q)$ from the PY theory.
Cross terms $(\ell \ne \ell')$ are omitted to avoid
overcrowding of the figures.
The function $S_{q}^{N}$ from the RISM theory and the one
based on Eq.~(\ref{eq:Sq-N-decom}) with the coefficients
$S_{\ell \ell'}^{0}(q)$ from the PY theory are found to be
in good agreement with each other, 
and therefore 
it makes sense to discuss the decomposition of $S_{q}^{N}$
using the results from two different integral-equation theories.
As mentioned in connection with Fig.~\ref{fig:Sq-near-phic},
the glass transition of our system is driven by the first
peak in $S_{q}^{N}$ centered at $q \approx 7$. 
Figure~\ref{fig:Sq-decom} shows that,
for the small elongation,
the first peak is primarily determined by the center-of-mass
density fluctuations, 
whilst the contributions from higher order angular correlations are
responsible only for the peaks located in the
higher-$q$ region.
On the other hand, when the elongation is large,
the contribution from the center-of-mass degrees of freedom gets
suppressed, but the higher order angular correlations become
much more important for determining the first peak:
the first peak is primarily accounted for by the 
$(2,2)$ contribution. 
Thus, the static density fluctuations determining the cage
for the glass transition 
are of different origins
for small and large elongations, respectively. 

A comment shall be added concerning the strong peak in 
$S_{22}^{0}(q)$ at $q \approx 0$ shown in 
Fig.~\ref{fig:Sq-llm} for $\zeta = 1.0$.
It is clear from Eq.~(\ref{eq:Sq-N-decom}) that the function
$S_{22}^{0}(q)$ contributes to $S_{q}^{N}$ with a prefactor
$j_{2}(q\zeta/2)^{2}$, which is proportional to $q^{4}$ for
small $q$. 
As a result, this strong peak in $S_{22}^{0}(q)$ hardly 
contributes to $S_{q}^{N}$ in the small wave-number regime;
it only gives rise to a small peak centered at
$q \approx 3.5$ as shown in Fig.~\ref{fig:Sq-decom}.
Also, it is seen that
the $(0,0)$ component has a peak at the same $q$ range.
However, it is found that the small peak at $q \approx 3.5$
is canceled out by the $(0,2)$ component, 
which is not shown in the figure.
All this together results 
in the small and flat $S_{q}^{N}$ for $q < 5$. 
We conclude that 
the strong peak in $S_{22}^{0}(q)$ at $q \approx 0$ 
is irrelevant for 
the glass formation for the elongation $\zeta = 1.0$
within our theory. 

The intramolecular correlation functions 
$w_{q}^{x}$ ($x = N,Z$) from Eq.~(\ref{eq:rho-NZ-d}) 
enter the mode-coupling vertices implicitly via the site-site
Ornstein-Zernike
equation for ${\bf S}_{q}$ and explicitly via Eq.~(\ref{eq:MCT-sNZ}). 
Using Eq.~(\ref{eq:wq-decom}), 
they can be decomposed in analogy to Eq.~(\ref{eq:Sq-N-decom}):
\be
w_{q}^{N \, (Z)} = 
\sum_{\ell \, : \, {\rm even} \, {\rm (odd)}}
2 (2\ell+1) \, j_{\ell}(q\zeta/2)^{2}.
\label{eq:wq-x-decom}
\ee
$w_{q}^{N \, (Z)}$ starts for $q = 0$ at the value 2 (0)
and then it oscillates for $q > 5$ around the value 1.
The first oscillation minima of $w_{q}^{N}$ occur at
$q = 11.4$ and 27.4 for $\zeta = 0.4$, and at
$q = 4.5$ and 11.0 for $\zeta = 1.0$.
For $w_{q}^{Z}$, the first minima are located at
$q = 19.4$ and 35.0 for $\zeta = 0.4$, and at $q = 7.8$ and 14.2
for $\zeta = 1.0$.

\begin{figure}
\includegraphics[width=0.7\linewidth]{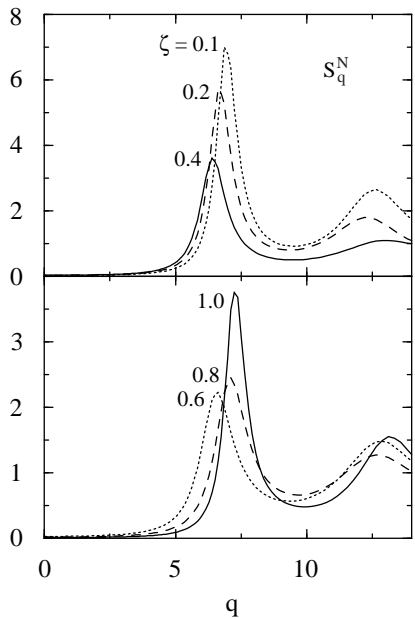}
\caption{
Static structure factors $S_{q}^{N}$ for the total density fluctuations
at the fixed packing fraction $\varphi = 0.56$ for various elongations
as indicated in the figure.}
\label{fig:Sq-N-vs-zeta}
\end{figure}

Let us consider the change of the structure factor
as function of the elongation $\zeta$ for fixed packing 
fraction $\varphi$.
Figure~\ref{fig:Sq-N-vs-zeta} exhibits the result for
$\varphi = 0.56$. 
It is seen that for small elongations (the upper panel)
the first peak height decreases with increasing the elongation,
whilst the opposite trend is seen for large elongations
(the lower panel).
This feature can be explained in terms of the spherical-harmonic
expansion coefficients $S_{\ell \ell}^{0}(q)$ as follows. 
As discussed in connection with 
Fig.~\ref{fig:Sq-decom},
the center-of-mass density fluctuations ($\ell = 0$) 
are primarily responsible for
determining the first peak in $S_{q}^{N}$ for small elongations,
whereas the angular correlations of the index $\ell = 2$
are more relevant for large elongations.
The strength of the center-of-mass 
correlation becomes weaker as the elongation is increased,
and this explains the decrease of the first peak height in $S_{q}^{N}$
for small elongations. 
On the other hand, when the elongation is large, $\ell = 2$ component
is relevant, and this angular correlation 
becomes stronger with increasing the elongation.
This explains the increase of the first peak height in $S_{q}^{N}$
with increasing $\zeta$ for large elongations.
Thus, the non-monotonic $\zeta$-dependence of the first peak height in
$S_{q}^{N}$ for the fixed packing fraction shown in 
Fig.~\ref{fig:Sq-N-vs-zeta} is due to the 
different origin of that peak for small and large elongations. 

\subsection{Preferred orientations for nearest neighbors}
\label{subsec:orientations}

A digression might be adequate for a better understanding
of the equilibrium structure of our molecular systems, 
especially, of angular correlations for nearest neighbors. 
Such angular correlations can best be investigated through the
molecular pair correlation function
$g(r_{12},\theta_{1},\theta_{2},\phi_{12})$. 
Here $r_{12}$ denotes the center-to-center separation,
and the three angles $\theta_{1}$, $\theta_{2}$,
and $\phi_{12} \equiv \phi_{1}-\phi_{2}$ specify the relative orientations
of the two linear molecules in the so-called $r$-frame. 
The pair correlation function can be expanded as~\cite{Gray84}
\bea
g(r_{12},\theta_{1},\theta_{2},\phi_{12}) &=& 
4\pi
\sum_{\ell_{1},\ell_{2},m}
g_{\ell_{1} \ell_{2}}^{m}(r_{12}) 
\nonumber \\
& &
\,\,
\times 
Y_{\ell_{1}}^{m}(\theta_{1},\phi_{1}) \,
Y_{\ell_{2}}^{m}(\theta_{2},\phi_{2})^{*}.
\label{eq:gr-expand}
\eea
The $g_{\ell_{1} \ell_{2}}^{m}(r_{12})$ are the spherical harmonic expansion
coefficients, and can be calculated within the PY 
theory~\cite{Chen71,Gray84}.
In the present work, the coefficients are 
calculated up to cutoff $\ell_{\rm cut} = 6$.

\begin{figure}
\includegraphics[width=0.7\linewidth]{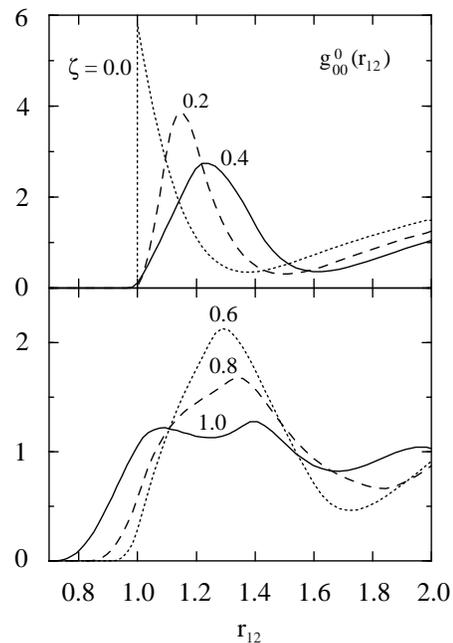}
\caption{
Center-of-mass component of the molecular pair correlation function,
$g_{00}^{0}(r_{12})$, as function of the center-to-center distance
$r_{12}$ 
at the fixed packing fraction $\varphi = 0.56$ 
for various elongations $\zeta$.}
\label{fig:gr-COM}
\end{figure}

The center-to-center radial distribution functions $g_{00}^{0}(r_{12})$
for representative elongations at the large packing fraction
$\varphi = 0.56$ 
are shown in Fig.~\ref{fig:gr-COM}, along with the 
radial distribution function for
hard spheres ($\zeta = 0.0$) at the same packing fraction.
As the elongation increases,
the first peak position in $g_{00}^{0}(r_{12})$ increases,
the height of the peak decreases, 
and the peak becomes broader
and somewhat irregular, with a shoulder developing at separations
just beyond $r_{12} = 1$. 
For the elongation $\zeta = 1.0$, the shoulder turns 
into a broad prepeak centered at $r_{12} \approx 1.1$. 

\begin{figure}
\includegraphics[width=0.7\linewidth]{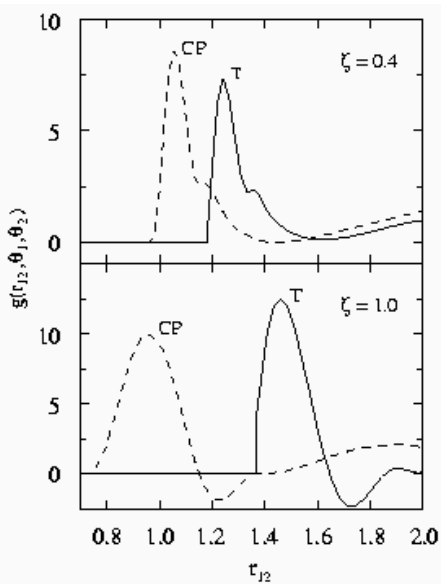}
\caption{
The $\phi_{12}$-averaged molecular pair correlation function
$g(r_{12},\theta_{1},\theta_{2})$ as defined in Eq.~(\ref{eq:gr-CP})
for the ``T-shaped'' ($\theta_{1} = 0$, $\theta_{2} = \pi/2$)
and the ``CP-type'' ($\theta_{1} = \theta_{2} = \pi/2$)
orientations (see text) 
at the packing fraction $\varphi = 0.56$ 
for the elongations
$\zeta = 0.4$ (upper panel) and $1.0$ (lower panel).}
\label{fig:gr-TCP}
\end{figure}

It becomes more difficult to interpret the
$g_{\ell_{1} \ell_{2}}^{m}(r_{12})$ for non-zero values of
$\ell_{1}$ and $\ell_{2}$. 
Therefore it seems adequate to follow
Streett and Tildesley~\cite{Streett76}
and consider cuts through the space of the four variables
determining the function $g$ in Eq.~(\ref{eq:gr-expand}).
Typical cuts for discussing the
relative orientations of two linear molecules are~\cite{Streett76}:
(i) the ``T-shaped'' orientation
($\theta_{1} = 0$, $\theta_{2} = \pi/2$, $\phi_{12} = \mbox{any value}$), 
(ii) the ``crossed'' orientation
($\theta_{1} = \theta_{2} = \phi_{12} = \pi/2$), 
(iii) the ``parallel'' orientation
($\theta_{1} = \theta_{2} = \pi/2$, $\phi_{12} = 0$), and
(iv) the ``end-to-end'' orientation
($\theta_{1} = \theta_{2} = 0$, $\phi_{12} = \mbox{any value}$).
These orientations lead to efficient packing at
close approach in the sense that they all lead to 
the close contact of the constituent atoms,
and thus contribute to the first peak in $S_{q}^{N}$.
Most of the orientations at high densities
can be broadly classified as 
being similar to one of these four.
Because of the computational reason to be described in the next
paragraph, the ``crossed'' and ``parallel'' orientations shall
be combined to define the ``CP-type'' orientation
($\theta_{1} = \theta_{2} = \pi/2$)
by averaging over the angle $\phi_{12}$:
\be
g(r_{12}, \theta_{1}, \theta_{2}) \equiv
\frac{1}{2\pi} \int_{0}^{2 \pi} d\phi_{12} \,
g(r_{12}, \theta_{1}, \theta_{2}, \phi_{12}).
\label{eq:gr-CP}
\ee
Notice that
the corresponding $\phi_{12}$-averaged pair correlation functions
for the ``T-shaped'' ($\theta_{1} = 0$, $\theta_{2} = \pi/2$)
and ``end-to-end'' ($\theta_{1} = \theta_{2} = 0$)
orientations remain 
the same as the original ones
since the angle $\phi_{12}$ is irrelevant in defining these
two orientations. 
Figure~\ref{fig:gr-TCP} exhibits representative results.

Before embarking on the conclusions to be drawn from 
Figs.~\ref{fig:gr-COM} and \ref{fig:gr-TCP}, 
let us make comments on the cutoff problem in the summation
in Eq.~(\ref{eq:gr-expand}).
To check the convergence, 
we also performed the same calculations with $\ell_{\rm cut} = 4$.
It is found that
for distances $r_{12} > 1 + \zeta$, the series converges rapidly,
i.e., the difference between the results with $\ell_{\rm cut} = 4$
and those with $\ell_{\rm cut} = 6$ is small. 
However, for distances $r_{12} < 1 + \zeta$, the 
difference is rather large reflecting the slow convergence of the
series, and this effect becomes more pronounced with increasing
density and elongation.
An indication of the lack of convergence is that
the functions for $\zeta = 1.0$ shown in the 
lower panel of Fig.~\ref{fig:gr-TCP} can
take unphysical negative values in the range
$r_{12} < 1 + \zeta$. 
The reason for the slow convergence in that range
is that the function $g(r_{12},\theta_{1},\theta_{2},\phi_{12})$
has step-like features because of the hard-core repulsion,
which cannot be accurately 
represented by a truncated series with a small
number for $\ell_{\rm cut}$. 
This problem is likely to be a feature of the spherical harmonic
expansion for any model in which the molecule has a relatively
hard asymmetric core.
It is also found that a better convergence is
achieved for the $\phi_{12}$-averaged correlation function defined in 
Eq.~(\ref{eq:gr-CP}) than the original 
$g(r_{12},\theta_{1},\theta_{2},\phi_{12})$,
and this is why we have chosen the averaged ones to display
the results. 
Despite these unwelcome features, 
it is anticipated that qualitative features of the angular
correlations are captured even with $\ell_{\rm cut} = 6$. 
Notice that the mentioned cutoff problem do not influence
the results to be presented for the MCT
since those are based solely on $S_{q}^{N}$ 
calculated from the RISM theory. 

The increase of the most probable nearest neighbor center-to-center
separation with increasing elongation, which is demonstrated
in Fig.~\ref{fig:gr-COM},  
suggests that the majority of nearest neighbor 
pairs adopt orientations for which the center-to-center distance of
closest approach increases with increasing elongation.
Therefore, it seems likely that ``CP-type'' orientations
do not contribute heavily to this peak in $g_{00}^{0}(r_{12})$, 
because their closest approach remains in the region
$r_{12} \approx 1$ irrespective of the elongation.
It is also clear that ``end-to-end'' orientations are unimportant,
because their minimum approach distance ($r_{12} = 1 + \zeta$)
lies well beyond the distance at which the first maximum occurs
in $g_{00}^{0}(r_{12})$.
Hence, the major contributions to the first peak in $g_{00}^{0}(r_{12})$ 
are likely to come from orientations of 
the ``T-shaped'' type and ones close to it.
The first maximum in the $g(r_{12},\theta_{1},\theta_{2})$
for the ``T-shaped'' orientation 
for $\zeta = 0.2$, 0.4, 0.6, 0.8, and 1.0 occur at
$r_{12} = 1.13$, 1.24, 1.33, 1.41, and 1.46, respectively,
as shown in Fig.~\ref{fig:gr-TCP} for $\zeta = 0.4$ and $1.0$. 
These positions are very close to the
first maximum positions in $g_{00}^{0}(r_{12})$ for each elongation
shown in Fig.~\ref{fig:gr-COM}.
This evidence is consistent with a strong predominance of 
``T-shaped'' nearest neighbor orientations.
We therefore conclude that at the most probable nearest neighbor
distance there is a strong preference for ``T-shaped'' orientations
over all others.

We next consider how the ``CP-type'' correlations manifest
themselves in the $g_{00}^{0}(r_{12})$.
As can be inferred from Fig.~\ref{fig:gr-TCP},
such correlations would lead to a peak centered at 
$r_{12} \approx 1$ irrespective of the elongation. 
This contribution leads to a small shoulder as shown in
Fig.~\ref{fig:gr-COM} for $\zeta = 0.6$. 
For the larger elongation $\zeta = 0.8$, 
the shoulder gets more pronounced,
and subsequently it leads to a broad peak at $r_{12} \approx 1.1$
for $\zeta = 1.0$.

Unlike hard spheres for which two centers cannot approach closer
than $r_{12} = 1$, two hard dumbbells can 
reach center-to-center distance $r_{12} < 1$
by adopting a ``crossed'' orientation.  
This explains why the $g_{00}^{0}(r_{12})$ in Fig.~\ref{fig:gr-COM}
are positive even for $r_{12} < 1$.
The probability of the ``crossed'' orientation increases with increasing
density because it relieves the strain of the closely packed system.
This effect is more pronounced at high elongations 
as shown in Fig.~\ref{fig:gr-TCP}, and is 
a major factor contributing to the growth of the shoulder in 
$g_{00}^{0}(r_{12})$
for the region $r_{12} < 1$ with increasing the elongation.

Let us add one final comment.
It is found from the extensions of 
the lower panel of Fig.~\ref{fig:gr-TCP}
to larger $r_{12}$ region that
both the ``T-shaped'' and ``CP-type''
correlations are rather long-ranged.
This is a manifestation of the strong peak in $S_{22}^{0}(q)$
at $q \approx 0$ shown in Fig.~\ref{fig:Sq-llm} for 
$\zeta = 1.0$,
as was discussed also in 
Refs.~\onlinecite{Letz00} and \onlinecite{Theenhaus01}.
The oscillatory feature of the above mentioned angular correlations
are found to continue up to 
$r_{12} \approx 2\pi / \Delta q$, where $\Delta q$ denotes the half width
of that peak in $S_{22}^{0}(q)$.
This intermediate-range order is absent in the case
of small elongations, say $\zeta = 0.4$.

\subsection{Bonding effects}
\label{subsec:bonding}

\begin{figure}
\includegraphics[width=0.7\linewidth]{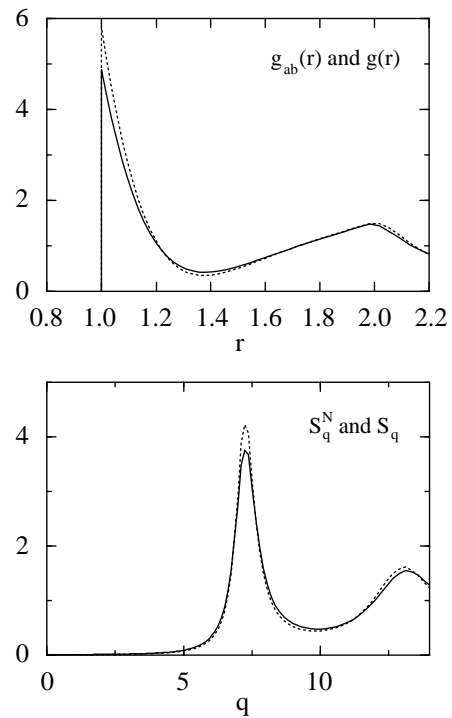}
\caption{
Upper panel: the site-site radial-distribution function 
$g_{ab}(r)$ 
for the symmetric-hard-dumbbell system (see text) with $\zeta = 1.0$ 
at the packing fraction $\varphi = 0.56$ (solid line), and
the radial-distribution function $g(r)$ for the hard-sphere system ($\zeta = 0.0$)
at the same packing fraction (dotted line).
Lower panel: the corresponding static structure factor $S_{q}^{N}$ for the total 
density fluctuations 
for the symmetric-hard-dumbbell system,
and the static structure factor $S_{q}$ for 
the hard-sphere system.}
\label{fig:gr-Sq-bonding}
\end{figure}

The dumbbell liquid for $\zeta = 1.0$ can be viewed as a system
of hard spheres of diameter $d = 1$ whose density is $2 \rho$
and where some additional covalent interaction has forced pairs
to be formed.
Let us consider the difference between the hard-sphere system and
the bonded system in detail. 
Figure~\ref{fig:gr-Sq-bonding} compares the 
the radial-distribution functions and the static structure factors
for the two systems 
at the fixed packing fraction $\varphi = 0.56$.  
Notice that the site-site radial-distribution
function $g_{ab}(r)$ for the symmetric dumbbell system becomes
independent of the site indices $a$ and $b$, and this is the 
adequate quantity to be compared with the radial-distribution 
function $g(r)$ for the hard-sphere system.
On the other hand, 
the total density static structure factor $S_{q}^{N}$ for 
the dumbbell system is the relevant one
to be compared with the static structure factor $S_{q}$ for the
hard-sphere system.
This is because,
when the packing fraction is fixed, the number density for 
the hard dumbbells with $\zeta = 1.0$ 
is half of that for the hard spheres, 
and the function $S_{q}^{N}$ properly accounts for this difference.
The functions $S_{q}^{N}$ and $S_{q}$ are also the relevant
inputs for the MCT equations for the hard-dumbbell and hard-sphere systems,
respectively. 

It is seen from the upper panel of Fig.~\ref{fig:gr-Sq-bonding} 
that the agreement of the radial-distribution functions for the two systems is
very good except for the first-coordination-shell region.
To demonstrate that this difference is primarily
caused by the bonding,
we shall consider the coordination number $K$, 
i.e., the number of nearest neighboring
spheres surrounding a central sphere, which can be calculated from the 
radial-distribution function.
It is found for the hard-sphere system that
$K = 12.1$ at $\varphi = 0.56$,
which is a typical value for simple systems at high density.
So, $K$ would tend to a value 12
also for the molecular system 
if the second sphere in one molecule was not
attached to the first sphere. 
However, we found 
$K = 11.4$ for $\zeta = 1.0$ at $\varphi = 0.56$. 
Thus, as should be expected, the second sphere in one molecule 
excludes one sphere in another
from being a nearest neighbor to the first sphere,
and this results in the reduction of the radial-distribution
function in the first shell region as shown in the upper panel.

The found feature for the radial-distribution functions also
explains the reduction of the first peak height in the static structure 
factor for $\zeta = 1.0$ compared to that for $\zeta = 0.0$, 
as exhibited in the lower panel of Fig.~\ref{fig:gr-Sq-bonding}.
Let us consider what would happen to the static structure
factor for the hard-sphere system when a short-ranged 
attractive force is added. 
This problem has been discussed for a square-well system~\cite{Dawson00}.
As demonstrated there, 
the attraction causes bonding, in the sense that the most probable
separation of two particles is smaller than expected for a 
pure hard-sphere system. 
This leads to the shift of the first peak position in the
static structure factor to higher $q$, 
the decrease of the peak height, and the increase of the peak 
wings~\cite{Dawson00}.
Although the first feature is not so prominent, 
the static structure factor for the $\zeta = 1.0$ dumbbell 
molecules reflects 
these features when compared to that for the hard-sphere system. 

We conclude that the structure of the cage for 
the hard-dumbbell system with $\zeta = 1.0$ 
is very close to the one for the hard-sphere system,
and that the difference can be explained 
as being due to the bonding effect.

\section{Structural Arrest}
\label{sec:4}

\subsection{Critical nonergodicity parameters}
\label{subsec:fq}

\begin{figure}
\includegraphics[width=0.7\linewidth]{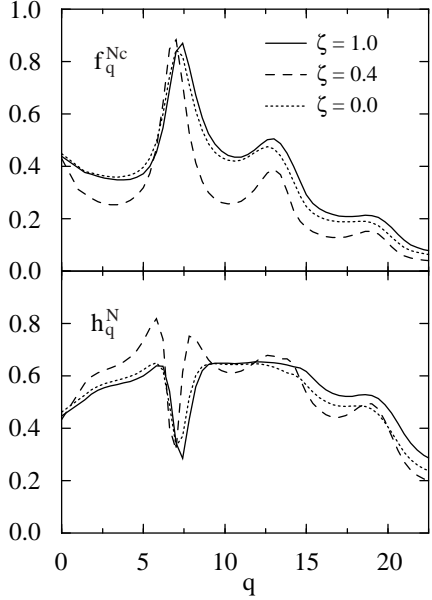}
\caption{
Critical nonergodicity parameter $f_{q}^{Nc}$ 
and critical amplitude $h_{q}^{N}$ for the coherent total density fluctuations
for the elongation $\zeta = 1.0$ (full lines), $\zeta = 0.4$ (dashed lines),
and for the hard-sphere system ($\zeta = 0.0$, dotted lines).}
\label{fig:fq-hq}
\end{figure}

The upper panel of Fig.~\ref{fig:fq-hq}
exhibits the results for the nonergodicity parameters
$f_{q}^{Nc}$ at the critical point $\varphi = \varphi_{c}$
for the elongations $\zeta = 0.4$ and $1.0$
calculated from Eq.~(\ref{eq:DW-N}). 
These are Debye-Waller factors of the system.
They can be measured, in principle, as cross section for
coherent neutron scattering. 
For large $S_{q}^{N}$, the compressibility 
$\kappa_{q} \propto S_{q}^{N}$ is large.
Therefore, spontaneous arrest is easier for larger $S_{q}^{N}$, 
and $f_{q}^{Nc}$ exhibits a maximum near the first peak position 
of $S_{q}^{N}$.
With varying $q$, $f_{q}^{Nc}$ oscillates in phase with 
$S_{q}^{N}$ ({\em cf.} Fig.~\ref{fig:Sq-near-phic}).
If the packing fraction increases, the arrested glass structure
stiffens, i.e. the
$f_{q}^{N}$ increases.
Expanding this increase for small distance parameters
$\epsilon = (\varphi - \varphi_{c})$, 
one finds~\cite{Franosch97,Fuchs98}
\be
f_{q}^{N} = f_{q}^{Nc} + D \sqrt{(\varphi - \varphi_{c})} \, h_{q}^{N} +
O(\varphi - \varphi_{c}).
\label{eq:fq-N-near-phic}
\ee
The critical amplitude $h_{q}^{N}$ is positive.
It characterizes the susceptibility of the arrested structure with 
respect to changes of the control parameters.
The formulas for the evaluation of $h_{q}^{N}$ and of the
constant $D > 0$ will be considered in the 
subsequent paper~\cite{Chong-MCT-dumbbell-2}.
Since $f_{q}^{N} \le 1$,
$f_{q}^{N} - f_{q}^{Nc}$ is bounded by $1-f_{q}^{Nc}$.
Therefore, the critical amplitude $h_{q}^{N}$ for the increase
of $f_{q}^{N}$
is much smaller for $q \approx 7$ than for $q$
off the structure-factor-peak position,
as shown in the lower panel of 
Fig.~\ref{fig:fq-hq}. 
These features are 
analogous to those found in the hard-sphere
system~\cite{Bengtzelius84,Franosch97}.

\begin{figure}
\includegraphics[width=0.7\linewidth]{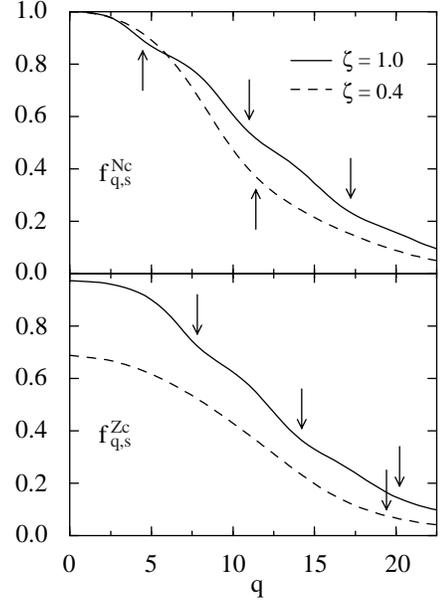}
\caption{
Critical Lamb-M\"ossbauer factors 
for the tagged molecule's total density fluctuations $f_{q,s}^{Nc}$ (upper panel)
and for the charge-density fluctuations 
$f_{q,s}^{Zc}$ (lower panel) 
for the elongation $\zeta = 1.0$ (full lines) and 
$\zeta = 0.4$ (dashed lines).
The arrows indicate the wave numbers for the minimum positions in 
$w_{q}^{N}$ and $w_{q}^{Z}$, respectively,
calculated from Eq.~(\protect\ref{eq:wq-x-decom}).}
\label{fig:fqs}
\end{figure}

Figure~\ref{fig:fqs} exhibits 
the tagged-molecule's critical nonergodicity parameters 
$f_{q,s}^{xc}$ ($x = N$, $Z$) 
for the elongations $\zeta = 1.0$ and $0.4$
calculated from Eq.~(\ref{eq:DW-sNZ}). 
These are Lamb-M\"ossbauer factors describing the arrested 
probability distribution of the tagged molecule. 
As expected for a 
localized-distribution Fourier transform, the
$f_{q,s}^{xc}$--versus--$q$ curves decrease with increasing $q$. 
The critical Lamb-M\"ossbauer factor $f_{q,s}^{Nc}$
for the total number density fluctuations 
approaches unity for $q$ tending to zero
due to the particle-number conservation law.
On the other hand, there is no analogous conservation
law for $x = Z$, and therefore one gets
$f_{q \to 0,s}^{Zc} < 1$. 

A remarkable feature of Fig.~\ref{fig:fqs} are
the gentle oscillations exhibited by $f_{q,s}^{Nc}$ and $f_{q,s}^{Zc}$.
In analogy to the discussion for $f_{q}^{Nc}$, 
it is expected that $f_{q,s}^{xc}$ oscillates 
in phase with $w_{q}^{x}$.
As discussed in connection with Eq.~(\ref{eq:wq-x-decom}), 
the function $w_{q}^{x}$ exhibits minima because of various
$j_{\ell}(q\zeta/2)^{2}$ contributions,
and these minimum positions are marked as arrows in
Fig.~\ref{fig:fqs}. 
One finds that the positions of the oscillations are well reproduced
by the arrows, indicating that they are due to the
presence of various angular-momentum-index $\ell$ contributions
to intramolecular interference effects.
A more definitive analysis concerning the origin of the oscillations
should be based on the decomposition of $f_{q,s}^{xc}$
in terms of the nonergodicity parameters of the 
tensorial density correlators 
$\Phi_{s,\ell \ell'}^{0}(q,t)$
introduced in connection with Eq.~(\ref{eq:Fqs-decom})~\cite{Chong01}.
Under the diagonal approximation
$\Phi_{s,\ell \ell'}^{0}(q,t) \approx \delta_{\ell \ell'} \Phi_{s,\ell \ell}^{0}(q,t)$,
one gets from the long-time limit of Eq.~(\ref{eq:Fqs-decom}):
\be
f_{q,s}^{xc} =
(2/w_{q}^{x}) \sum_{\ell}
(2 \ell + 1) j_{\ell}(q\zeta/2)^{2} f_{s}^{c}(q, \ell, 0).
\label{eq:fqs-x-decom}
\ee
Here, 
$f_{s}^{c}(q, \ell, 0) = \lim_{t \to \infty} \Phi_{s,\ell \ell'}^{0}(q,t)$
for $\varphi = \varphi_{c}$,
and $\ell$ should be even (odd) for $x = N$ ($Z$). 
As discussed in Ref.~\onlinecite{Chong01}, the gentle oscillations
exhibited by $f_{q,s}^{xc}$ 
can be explained as being due to interference effects of the
$f_{s}^{c}(q, \ell, 0)$ with the intramolecular form factors
$j_{\ell}(q\zeta/2)^{2}$.
Unfortunately, we do not have information on the 
$f_{s}^{c}(q, \ell, 0)$.

\begin{figure}
\includegraphics[width=0.7\linewidth]{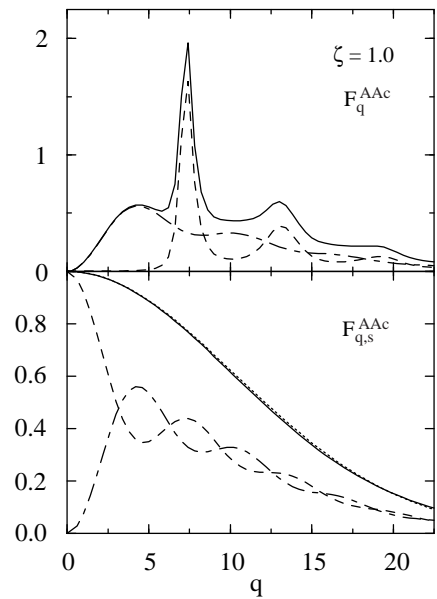}
\caption{
The full lines exhibit the 
critical nonergodicity parameters for the atomic density fluctuations
$F_{q}^{AAc}$ (upper panel) and its self part
$F_{q,s}^{AAc}$ (lower panel) 
for the elongation $\zeta = 1.0$.
The dashed and dashed-dotted lines denote contributions from the
total-density ($N$) and charge-density ($Z$) components, 
respectively. 
The dotted line in the lower panel denotes the 
result based on the Gaussian approximation for $F_{q,s}^{AAc}$, 
Eq.~(\ref{eq:Fqs-AA-Gaussian}).}
\label{fig:Fq-site-strong}
\end{figure}

It might be interesting to consider how the results for the total
density and charge density fluctuations can be translated to those
for the atomic density fluctuations.
The latter can be obtained from the former via the inverse
relation of Eqs.~(\ref{eq:phi-NZ}) and (\ref{eq:phi-sNZ}):
\begin{subequations}
\bea
F_{q}^{AA} &=& \frac{1}{2} ( f_{q}^{N} S_{q}^{N} + f_{q,s}^{Z} w_{q}^{Z}), 
\\
F_{q,s}^{AA} &=& \frac{1}{2} ( f_{q,s}^{N} w_{q}^{N} + f_{q,s}^{Z} w_{q}^{Z}).
\eea
\end{subequations}
Notice that the self part
$F_{q,s}^{AA}$ can be measured as cross section for incoherent
neutron scattering.
The results at the critical packing fraction 
are exhibited as solid lines in 
Fig.~\ref{fig:Fq-site-strong} for the elongation $\zeta = 1.0$.
The dashed and dashed-dotted lines are contributions from the
total density ($N$) and charge density fluctuations ($Z$),
respectively. 
A small peak centered at $q \approx 4$
develops in $F_{q}^{AAc}$ 
due to the charge-density fluctuations. 
The dotted line in the lower panel for the self part $F_{q,s}^{AAc}$ denotes the 
result based on the Gaussian approximation:
\be
F_{q,s}^{AAc} \approx e^{- q^{2} (r_{A}^{c})^{2} }.
\label{eq:Fqs-AA-Gaussian}
\ee
Here, $r_{A}^{c}$ is the critical localization length 
for atom $A$, defined via
$\lim_{q \to 0} (1 - F_{q}^{AAc}) / q^{2} = (r_{A}^{c})^{2}$. 
It is seen that 
the constituent atom's critical Lamb-M\"ossbauer factor $F_{q,s}^{AAc}$ 
is well described by a Gaussian, 
in particular, it does not exhibit oscillations.
It is surprising that the sum of two non-Gaussian functions is almost
Gaussian.
The analogous results for $\zeta = 0.4$ are quite similar,
except the peak of $F_{q}^{AAc}$ for $q \approx 4$ is
suppressed. 

\subsection{Phase diagram}
\label{subsec:phase}

The phase diagram in Fig.~\ref{fig:phase} can be understood
as result of the control-parameter dependence of the structure
factors which were explained in Sec.~\ref{sec:3}. 
A prominent feature is the maximum of the 
$\varphi_{c}$--versus--$\zeta$ curve near $\zeta = 0.43$. 
This is because of the two different mechanisms 
for the structural arrest, one dominating 
for small and the other one for large elongations.
As discussed in connection with Fig.~\ref{fig:Sq-near-phic},
the glass transition is driven by the
first-peak region in the static structure factor $S_{q}^{N}$,
irrespective of the elongation.
For small elongations, the peak is primarily determined by the
center-of-mass density fluctuations, and its strength becomes
weaker with increasing the elongation as explained in 
connection with Fig.~\ref{fig:Sq-decom} and the upper
panel of Fig.~\ref{fig:Sq-N-vs-zeta}. 
Therefore, a relatively higher packing fraction is required 
to get into the glassy phase
if the elongation is increased, and this explains the increase of the
$\varphi_{c}(\zeta)$ curve for small elongations.
On the other hand, for large elongations, the first peak in $S_{q}^{N}$
is mainly determined by the $\ell = 2$ angular correlation, and its
magnitude gets larger with increasing the elongation as also explained
in connection with Fig.~\ref{fig:Sq-decom} and the lower 
panel of Fig.~\ref{fig:Sq-N-vs-zeta}. 
Thus, a relatively lower 
packing fraction is required for the glass formation
as the elongation is increased, and this explains the
decrease of the $\varphi_{c}(\zeta)$ curve for large elongations.
As a result of these two competing mechanisms for the glass formation, 
the transition line $\varphi_{c}(\zeta)$ exhibits a maximum. 

Another remarkable feature results from the 
structure-factor-peak reduction due to bonding 
which was explained above in connection with 
Fig.~\ref{fig:gr-Sq-bonding}.
This reduction stabilizes the liquid phase.
As a result, the critical packing fraction for elongation
$\zeta = 1$, $\varphi_{c}(\zeta = 1) \approx 0.56$, is larger
than the one for the transition of the hard-sphere system, 
$\varphi_{c}(\zeta = 0) \approx 0.53$.
Combined with the results discussed in the preceding paragraph, 
this implies that for all $0 < \zeta \le 1$
the critical packing fraction of the hard-dumbbell system
is larger than that of the hard-sphere system;
the liquid phase gets expanded due to the formation of 
molecules.
The increased-free-volume phenomenon 
due to the bond formation is consistent
with the result discussed for a square-well system~\cite{Dawson00}.

There are two alternatives for the glassy states, 
phases II and III in Fig.~\ref{fig:phase}, 
with respect to the
charge-density dynamics of the tagged molecule.
Phase II deals with states for sufficiently small $\zeta$. 
There is such small steric hindrance for a flip of the 
tagged molecule's axis
between the two energetically equivalent positions 
${\vec e}_{s}$ and $- {\vec e}_{s}$ that Eq.~(\ref{eq:DW-sNZ}) 
for $x=Z$ yields $f_{q,s}^{Z} = 0$. 
The dynamics of the charge fluctuations is ergodic. 
In particular, the dipole correlator relaxes to zero: 
$C_{1,s} (t \to \infty) = 0$
where
$C_{1,s} (t) = \langle {\vec e}_{s}(t) \cdot {\vec e}_{s}(0) \rangle =
\phi_{q = 0,s}^{Z}(t)$~\cite{Chong01}. 
For sufficiently large $\zeta$, on the other hand, the steric
hindrance for dipole reorientations becomes so effective, that also the
charge fluctuations behave nonergodically. 
In this case, Eq.~(\ref{eq:DW-sNZ}) for $x=Z$
yields a positive long-time limit,  
$0 < f_{q,s}^{Z} = \phi_{q,s}^{Z} (t \to \infty)$. 
In particular, dipole disturbances do not relax to zero: 
$C_{1,s} (t \to \infty) = f_{1,s} = f_{q=0,s}^{Z} > 0$. 
This phase III is a glass
with all structural disturbances exhibiting nonergodic motion.
In particular, the nonergodicity parameter
$f_{1,s}^{c}$ for, say, $\zeta \ge 0.6$ is as large as the
maximum of $f_{q}^{Nc}$, Fig.~\ref{fig:fq-hq}. 
The two phases II and III are separated by a curve $\varphi_{A}(\zeta)$,
where $\varphi_{A}(\zeta) \ge \varphi_{c}(\zeta)$. 
This curve is shown as the dashed line in Fig.~\ref{fig:phase}.
Since the steric hindrance for the molecule's flip motion increases
with increasing $\zeta$, one might expect that the curve 
$\varphi_{A}(\zeta)$ would monotonically decrease
with increasing $\zeta$. 
However, this is not the case.
What monotonically decreases with increasing $\zeta$ is
the difference $\varphi_{A}(\zeta) - \varphi_{c}(\zeta)$,
and the variation of $\varphi_{c}(\zeta)$ dominates that of
$\varphi_{A}(\zeta)$ for small $\zeta_{c} - \zeta$. 
The curve $\varphi_{A}(\zeta)$ terminates at the critical 
elongation $\zeta_{c}$: 
$\varphi_{A}(\zeta_{c}) = \varphi_{c}(\zeta_{c})$.
For our model, one finds
$\zeta_{c} = 0.345$, and its position is marked by an arrow in 
Fig.~\ref{fig:phase}.
The asymptotic laws for the transition from phase II to phase III
have earlier been described as the type-$A$ transition, 
as can be inferred from Ref.~\onlinecite{Franosch94} and the papers quoted
there. 
The square-root singularity of the Debye-Waller factor $f_{q}^{N}$,
Eq.~(\ref{eq:fq-N-near-phic}),
implies via Eq.~(\ref{eq:DW-sNZ}) that the two phase transition
lines do not merge transversally:
$\frac{d}{d\zeta} \varphi_{A}(\zeta_{c}) = \frac{d}{d\zeta} \varphi_{c}(\zeta_{c})$. 
All together, this explains the minimum of 
the $\varphi_{A}(\zeta)$--versus--$\zeta$ curve near $\zeta = 0.23$. 

\begin{figure}
\includegraphics[width=0.7\linewidth]{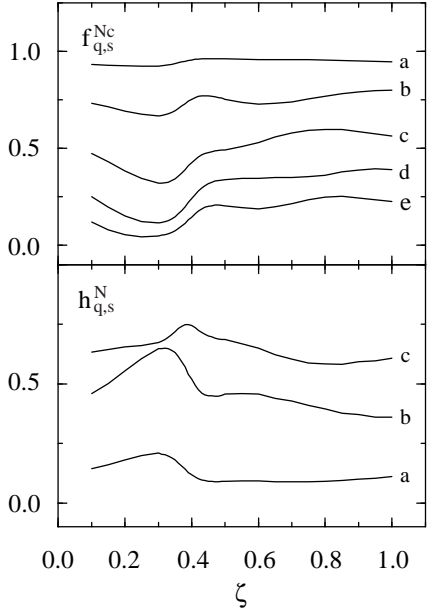}
\caption{
Critical nonergodicity parameters $f_{q,s}^{Nc}$ 
and the critical amplitudes $h_{q,s}^{N}$ for the 
tagged molecule's total density fluctuations 
along the type-$B$-transition line in Fig.~\protect\ref{fig:phase}
parameterized by the elongation $\zeta$
for the wave numbers $q = 3.4 (a), \, 7.0 (b), \, 10.6
(c), \, 14.2 (d)$, and 17.4 $(e)$.
The critical amplitudes $h_{q,s}^{N}$ for $q = 14.2$ and $17.4$ 
have been omitted to avoid the overlapping of the curves.}
\label{fig:fqs-N-vs-zeta}
\end{figure}

\begin{figure}
\includegraphics[width=0.7\linewidth]{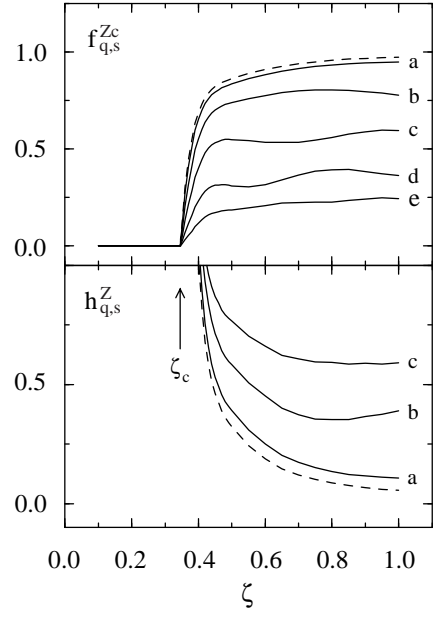}
\caption{
Results as in Fig.~\protect\ref{fig:fqs-N-vs-zeta}, but for the
tagged molecule's 
charge density fluctuations, $f_{q,s}^{Zc}$ and $h_{q,s}^{Z}$.
The dashed lines are added here, denoting the critical nonergodicity
parameter (upper panel) and the critical amplitude 
(lower panel) for the zero-wave-number limit. 
The arrow marks the transition point from phase II to phase III at 
$\zeta_{c} = 0.345$ taken from Fig.~\protect\ref{fig:phase}.}
\label{fig:fqs-Z-vs-zeta}
\end{figure}

There are some characteristic features of the type-$A$ transition
which are relevant in the analysis of 
the type-$B$-transition dynamics:
these are connected with the $\zeta$-variation of the critical 
Lamb-M\"ossbauer factors $f_{q,s}^{xc}$ and the critical amplitudes 
$h_{q,s}^{x}$~\cite{Chong-MCT-dumbbell-2}. 
Figure~\ref{fig:fqs-N-vs-zeta} exhibits 
$f_{q,s}^{Nc}$ and $h_{q,s}^{N}$ for the total density fluctuations
along the type-$B$ transition line $\varphi_{c}(\zeta)$ parameterized
by $\zeta$. 
Figure~\ref{fig:fqs-Z-vs-zeta} shows the corresponding results for
the charge-density correlator, $f_{q,s}^{Zc}$ and $h_{q,s}^{Z}$.
They deal with the transition from phase II for $\zeta < \zeta_{c}$
to phase III for $\zeta > \zeta_{c}$. 
We note in passing that the curves shown in these figures exhibit
non-monotonic $\zeta$ dependence, such as wiggles, 
or even minima and maxima.
These anomalies are analogues of the gentle oscillations, discussed above
in connection with Fig.~\ref{fig:fqs}, 
and they 
can in principle be explained as was done in Ref.~\onlinecite{Chong01}
for dumbbell molecules immersed in a hard-sphere system.
For strong steric
hindrance, say $\zeta \geq 0.8$, $f_{q,s}^{Nc}$ 
is rather close to
$f_{q,s}^{Zc}$, and this holds also for critical amplitudes,
$h_{q,s}^{N}$ and $h_{q,s}^{Z}$. 
For $\zeta$ approaching $\zeta_c$, the Lamb-M\"ossbauer factor
$f_{q,s}^{Zc}$ falls below
$f_{q,s}^{Nc}$, and the critical amplitude $h_{q,s}^{Z}$ grows 
above $h_{q,s}^{N}$.
These are characteristic features of the type-$A$ transition,
whose transition point can be characterized by the vanishing of the
critical nonergodicity parameter $f_{q,s}^{Zc}$ and by the
divergence of the critical amplitude $h_{q,s}^{Z}$~\cite{Franosch94}.
The former feature is demonstrated, for example, by the strong decrease of 
$f_{1,s}^{c} = f_{q \to 0,s}^{Zc}$ for $\zeta = 0.4$ 
shown in the lower panel of Fig.~\ref{fig:fqs} 
compared to the one for $\zeta = 1.0$ in the same panel. 
Since the critical amplitude
$h_{q,s}^{Z}$ gauges the dynamics in the $\beta$-relaxation
regime, 
the dynamics of the charge-density correlators
as well as the dipole correlator near $\zeta_{c}$
is strongly influenced by precursor phenomena of the type-$A$
transition from phase II to phase III.
Their dynamics in the $\alpha$-relaxation regime is also perturbed 
since the leading correction to the $\alpha$-scaling law is 
proportional to the critical amplitude~\cite{Franosch97}. 
These features will be discussed in the following
paper~\cite{Chong-MCT-dumbbell-2}.
Thus, the dynamics for elongations close to $\zeta_{c}$ 
is qualitatively different from the one for large elongations,
and this is why we have chosen
as the representative elongation $\zeta = 0.4$ 
for the demonstration of the state with weak steric hindrance
for reorientations.
Generically, there is no type-$A$ transition line for
arbitrary diatomic molecules. 
In our problem, this singularity is due to an additional symmetry
that produces vanishing coupling constants.
The top-down symmetry of the molecule renders the MCT equations
decouple completely into a set for total density fluctuations
and another one for charge-density fluctuations. 
For nearly top-down symmetrical molecules, the type-$A$ transition
is smeared to a rapid crossover from the very small
nonergodicity parameters $f_{q,s}^{Z}$ for $\zeta \ll \zeta_{c}$
to such of order unity for 
$\zeta \gg \zeta_{c}$~\cite{Franosch97c,Franosch94,Franosch98b}.
By continuity, the critical amplitude $h_{q,s}^{Z}$ remains 
large for $\zeta$ near $\zeta_{c}$ also for asymmetrical molecules.
Therefore, the results for $\zeta = 0.4$ 
are also representative for such cases, where the type-$A$
transition singularity is avoided due to breaking of the
top-down symmetry of the constituent molecules,
provided the breaking is sufficiently weak. 

The described type-$A$ transition has been studied also for 
a single dumbbell immersed in a system of hard spheres.
The diagrams corresponding to the upper panels in 
Figs.~\ref{fig:fqs-N-vs-zeta} and \ref{fig:fqs-Z-vs-zeta}
are qualitatively similar~\cite{Franosch98b},
but there are two remarkable differences.
First, the form factors $f_{q,s}^{Nc}$ are somewhat larger
and the variation with $\zeta$ for $\zeta \ge \zeta_{c}$
is more pronounced for the hard-dumbbell system 
than for the corresponding
quantities for the simple system.
Similarly, for $\zeta > 0.5$ the
$f_{q,s}^{Zc}$ are larger in Fig~\ref{fig:fqs-Z-vs-zeta}
than for the single-dumbbell system.
Second, the $(\zeta - \zeta_{c})$ interval for the decay
of $f_{q,s}^{Zc}$ from large weakly $\zeta$-dependent
values to zero at the transition is narrower for the
motion in the dumbbell liquid than for the motion in the
hard-sphere system.
These differences reflect the fact that steric hindrance
for translation as well as for reorientation is
more efficient if the cage-forming neighbor molecules
are sufficiently elongated rather than being spherical.
This conclusion explains also that the found critical value
$\zeta_{c} = 0.345$ is smaller than the corresponding 
value $0.380$ obtained for a single dumbbell in a
hard-sphere system~\cite{Chong01}. 

\section{Conclusions}
\label{sec:5}

A mode-coupling theory (MCT) for the evolution of glassy dynamics
is derived and used to discuss the idealized liquid-glass transition
in a hard-dumbbell system (HDS).
The theory predicts a singular change of the dynamics caused by
a regular change of the canonically defined equilibrium
structure factors with variations of control parameters like
the packing fraction $\varphi$.
The structure factors define the mode-coupling constants in the
equations of motion for the correlation functions,
and they have been evaluated within the RISM and Percus-Yevick
theories.
The good agreement of the results from the two approximate 
approaches support the opinion that the used input information
of the MCT is semiquantitatively correct.
The results have been used to demonstrate that
``T-shaped'' configurations are the preferred arrangements of
the cage-forming neighbors of a molecule.
The found arrangements are similar to those
discussed earlier for more dilute systems~\cite{Streett76};
but the ordering in our high-density regime is more pronounced.
In addition, there is intermediate-ranged order leading to a central peak for
quadrupole-density fluctuations, but this is irrelevant for
the explanation of the glassy dynamics 
within the present theory (Sec.~\ref{sec:3}).
There is no information available on the correctness of the cited 
structure-factor theories within the large-$\varphi$ regime studied
in this paper.
This implies obvious reservation concerning quantitative details
of the results presented.

The MCT for molecular systems proposed in this paper is based on 
describing the dynamics by $n$-by-$n$-matrix correlators
formed with the $n$ interaction-site densities of the
molecule's constituent atoms. 
Such basis is inferior to the one using a description by
infinite-matrix correlators formed with 
tensor-density fluctuations~\cite{Schilling97,Theis98,Fabbian99b,Winkler00,Theis00,Letz00,Theenhaus01,Franosch97c,Goetze00c},
provided the equations of motion of the latter theory could be
solved for parameters and time regimes of interest.
For example, our theory does not directly lead to results for the
angular-momentum-$\ell = 2$ reorientational correlator,
which is relevant for the description of depolarized 
light-scattering data.
However, it was shown already in some other context~\cite{Chong01},
how the $\ell = 2$ reorientational correlator can be obtained as an addendum
of the site-representation theory.
A more subtle extension of the theory would be necessary,
if there would be a second-order phase transition.
A treatment of the interference of the slow glass dynamics
with the critical dynamics of the phase transition would
require the inclusion of the critical fluctuations in the relaxation 
kernel in the spirit of the original derivation of 
mode-coupling theories by Kawasaki~\cite{Kawasaki66}.
It is unclear at present whether such an extension 
can be formulated.

Compared to a hard-sphere system (HSS),
the fusion of two hard spheres to a dumbbell of elongation $\zeta$,
$0 < \zeta \le 1$, increases the free volume
if the packing fraction is kept fixed.
Therefore, the liquid gets stabilized and the
line for the liquid-to-glass transition $\varphi_{c}(\zeta)$
is above the transition value $\varphi_{c}^{\rm HSS}$ of the HSS.
Like for the HSS, the transition is driven by the density fluctuations
with wave vectors near the position of the first sharp diffraction peak.
For small $\zeta$, the peak is formed by the center-to-center 
correlations which decrease with increasing $\zeta$, leading
to an increase of the $\varphi_{c}$--versus--$\zeta$ transition curve.
For large $\zeta$, the peak is formed by the quadrupole correlations;
and these increase with $\zeta$, leading to a decrease
of the transition line.
This explains the pronounced maximum of the transition curve in 
Fig.~\ref{fig:phase}.
The model studied exhibits a symmetry with respect to the
top-down flip of the molecule's axis.
As explained in the earlier MCT literature,
this implies a line of spin-glass-type transitions shown as
dashed curve in Fig.~\ref{fig:phase}.

A comment concerning the accuracy of the reported calculations might
be adequate.
After the specified discretization of the wave numbers, 
Eq.~(\ref{eq:DW-N}) for the 100 numbers $f_{q}^{N}$ is solved by the iteration
$f_{q}^{N \, (j+1)} = 
{\cal F}_{q}^{N}[f^{N \, (j)}] / \{ 1 + {\cal F}_{q}^{N}[f^{N \, (j)}] \}$,
$j = 0, 1, \cdots$, starting from $f_{q}^{N \, (0)} = 1$.
The sequence decreases monotonically towards the nonergodicity parameter
$f_{q}^{N \, (j)} \to f_{q}^{N}$.
The linearized iteration for 
$\delta f_{q}^{(j)} = f_{q}^{N \, (j)} - f_{q}^{N}$ reads
$\delta f_{q}^{(j+1)} = \sum_{p} A_{qp} \delta f_{p}^{(j)}$,
where the Frobenius matrix ${\bf A}$ is given by 
$A_{qp} = (1-f_{q}^{N})^{2} \partial {\cal F}_{q}^{N}[f^{N}] / \partial f_{p}^{N}$.
The matrix has a maximum eigenvalue $E \le 1$.
Off the critical points, one gets $E < 1$, and the convergence of the 
iteration is exponentially fast.
The critical point is characterized by $E^{c} = 1$, and here
the convergence is only algebraically.
The proofs of the cited mathematical properties can be found in
Ref.~\onlinecite{Goetze95b}.
Near the critical point, one derives from Eq.~(\ref{eq:fq-N-near-phic}): 
$E^{c} - E \propto \sqrt{\varphi - \varphi_{c}}$.
Analogous statements hold for the calculations of 
$f_{q,s}^{Z}$ from Eq.~(\ref{eq:DW-sNZ}). 
In our numerical work, the value for $E$ is controlled and
$\varphi - \varphi_{c}$ is determined routinely so, that
$E^{c} - E \approx 10^{-4}$.
Hence, the critical points are calculated with an accuracy 
of the order $10^{-8}$.
Thus, the accuracy of the lines in Fig.~\ref{fig:phase} is determined 
by the number $n^{*}$ of values for $\zeta$ used to calculate
$\varphi_{c}(\zeta)$ and $\varphi_{A}(\zeta)$.
We used a grid with $n^{*} = 70$;
it was chosen non-uniformly over the interval $0 \le \zeta \le 1$
with the highest density of points for $\zeta$ near $\zeta_{c}$.

Letz {\em et al.} have discussed a liquid-glass phase diagram
for a system of hard ellipsoids~\cite{Letz00}.
Considering their aspect ratio of the prolate ellipsoids 
as analogue of $(\zeta + 1)$ for the dumbbells, their 
phase diagram looks similar to Fig.~\ref{fig:phase}.
They also show the analogue of the glass-to-glass
transition curve, albeit without a minimum and with a 
transversal termination at the liquid-glass-transition line.
It is argued in Ref.~\onlinecite{Letz00} that the strong decrease
of the $\varphi_{c}$--versus--$\zeta$ curve for aspect ratios
near $2$ is an implication of the central peak of the 
angular-momentum-$\ell = 2$ correlations,
reflecting a nematic-transition precursor.
Hence, the explanation of the phase diagram given 
in Ref.~\onlinecite{Letz00}
for the hard-ellipsoid system is quite different from the explanation
of Fig.~\ref{fig:phase} for the dumbbell system.

The critical form factors for the glass $f_{q}^{Nc}$ quantify
the arrested amorphous density fluctuations at the transition.
The wave-vector dependence, Fig.~\ref{fig:fq-hq}, 
is quite similar to that for a HSS.
This reflects the fact that the cage around an
interaction site is quite similar to the one found for the HSS.
The critical nonergodicity parameters $f_{q,s}^{Zc}$
for the arrest of the dipole reorientation, 
Fig.~\ref{fig:fqs-Z-vs-zeta},
are larger than the same quantities
calculated for a single molecule in the HSS~\cite{Chong01}.
In particular, the decrease of $f_{q=0,s}^{Zc}$ for $\zeta$
decreasing to the critical value $\zeta_{c}$ is so abrupt,
that the transition looks similar to a discontinuous one.
This shows that steric hindrance for reorientation
is more effective in a molecular system than in a system of
spherical particles.

In the following paper~\cite{Chong-MCT-dumbbell-2},
it will be shown that the results for the arrested
structure provide the key for an explanation of the 
structural relaxation.

\begin{acknowledgments}
We thank cordially W.~Kob, R.~Schilling, M.~Sperl, and Th.~Voigtmann
for constructive critique of our manuscript.

\end{acknowledgments}

\appendix

\section{Mode-coupling theory for molecular systems}
\label{appen:A}

The MCT focuses on the dynamics of density fluctuations.
Within the site representation, the basic variables are
the density fluctuations for the $n$ interaction sites
of the $N$ molecules:
$\rho_{\vec q}^{a} = \sum_{i=1}^{N} \exp( i {\vec q} \cdot {\vec r}_{i}^{\, a})$,
$a = 1,2,\cdots,n$.
Here ${\vec r}_{i}^{\, a}$ denotes the position vector of the site $a$ in 
molecule $i$. 
The most important correlation functions for a statistical description
of the dynamics are
\begin{equation}
F^{ab}_{q}(t) = 
\langle \rho^{a}_{\vec q}(t)^{*} \rho^{b}_{\vec q}(0) \rangle / N, \quad
a,b = 1, \cdots, n.
\end{equation}
These $n^{2}$ functions shall be considered as the elements of an 
$n \times n$ matrix ${\bf F}_{q}(t)$.
This matrix is real and symmetric.
The short-time expansion of this matrix is given by
\begin{equation}
{\bf F}_{q}(t) = {\bf S}_{q} - 
{\textstyle \frac{1}{2}} \, q^{2} \, {\bf J}_{q} \, t^{2} +
{\bf O}(t^{4}).
\end{equation}
Here the structure-factor matrix is given by
$S_{q}^{ab} = \langle \rho^{a}_{\vec q}(t)^{*} \rho^{b}_{\vec q}(0) \rangle / N$,
and 
$J_q^{ab} = 
\langle 
({\vec q} \cdot {\vec j}_{\vec q}^{\, a})^* 
({\vec q} \cdot {\vec j}_{\vec q}^{\, b}) 
\rangle / N q^2$
is defined in terms of the currents referring to the interaction sites
$\vec j_{\vec q}^a = \sum_{i=1}^{N} {\vec v}_{i}^{\, a} 
\exp(i {\vec q} \cdot {\vec r}_{i}^{\, a})$
with ${\vec v}_{i}^{\, a}$ denoting the velocity of the site $a$ in molecule $i$.
The Zwanzig-Mori formalism~\cite{Hansen86} leads to an 
exact equation of motion for ${\bf F}_{q}(t)$:
\begin{equation}
\partial_{t}^{2} {\bf F}_{q}(t) + {\bf \Omega}_{q}^{2} \, {\bf F}_{q}(t) +
{\bf \Omega}_{q}^{2} 
\int_{0}^{t} dt' \, {\bf m}_{q}(t-t') \, \partial_{t'} {\bf F}_{q}(t') = {\bf 0},
\label{eq:GLE-v-a}
\end{equation}
where
\begin{equation}
{\bf \Omega}_{q}^{2} = q^{2} \, {\bf J}_{q} \, {\bf S}^{-1}_{q}.
\label{eq:GLE-v-b}
\end{equation}
The right-hand side of this equation is a product of two positive
definite matrices. 
Hence it is equivalent to the square of a positive definite matrix.
Therefore, one can write it as the square, ${\bf \Omega}_{q}^{2}$,
of some matrix ${\bf \Omega}_{q}$. 
Splitting off this matrix in front of the
convolution integral is done for later convenience.

The difficult problem is the derivation of an approximation for
the matrix ${\bf m}_{q}(t)$ of fluctuating-force correlations 
such that the
cage effect is treated reasonably. 
This has been done originally in Ref.~\onlinecite{Chong98b}
by extending the procedure used for atomic systems~\cite{Goetze91b}.
But the reported formulas~\cite{Chong98b} are not acceptable.
Firstly, they do not properly reduce to the ones for
simple systems in the united atom limit.
Secondly, the momentum conservation law for coherent
dynamics is not satisfied.
For these reasons, an alternative derivation has been developed 
in Ref.~\onlinecite{Chong01} starting from the projection-operator
theory of Mori and Fujisaka~\cite{Mori73},
albeit for molecules immersed in a simple system.
It is possible to generalize this derivation 
for the coherent density correlators 
${\bf F}_{q}(t)$ for molecular systems, but the procedure becomes
more involved; it shall be described in a separate 
paper~\cite{Chong-MCT-deri}. 
Here, a more simplified derivation shall be presented.

The simplified procedure starts by assuming that a molecular system
is a mixture of constituent atoms;
intramolecular constraints between constituent atoms are 
accounted for by the pair correlations only. 
In this way, a system of $N$ molecules is treated as a mixture
of $n$ species, each consisting of $N$ particles. 
Using the equations in MCT for mixtures~\cite{Barrat90b}, 
one gets the following 
expression for the relaxation kernel:
\begin{equation}
m^{ab}_{q}(t) = {\cal F}^{ab}_{q}[{\bf F}(t)],
\label{eq:MCT-v-a}
\end{equation}
where the mode-coupling functional 
$\mbox{\boldmath ${\cal F}$}_{q}$ is given by
the equilibrium quantities:
\bea
& &
{\cal F}^{ab}_{q}[\, \tilde{\textit {\textbf f}} \, ] =
\frac{1}{2}  
\sum_{c} S^{ac}_{q} \, 
\sum_{\lambda, \lambda', \mu, \mu'}
\int d{\vec k} \, 
V_{\lambda \lambda' \mu \mu'}^{cb}({\vec q}; {\vec k}, {\vec p} \,) \,
\nonumber \\
& &
\qquad \qquad \qquad \qquad \qquad \qquad \qquad \quad
\times \, 
\tilde{f}_{k}^{\lambda \lambda'} \,
\tilde{f}_{p}^{\mu \mu'},
\label{eq:MCT-v-b}
\\
& &
V_{\lambda \lambda' \mu \mu'}^{ab}({\vec q}; {\vec k}, {\vec p} \,) =
\frac{\rho}{(2 \pi)^{3}} \, 
\{ {\vec q} \cdot [\delta^{a \mu}      \, {\vec k} \, c_{k}^{a \lambda} +
                   \delta^{a \lambda}  \, {\vec p} \, c_{p}^{a \mu}       ] \} 
\nonumber \\
& &
\qquad \qquad \quad
\times \,
\{ {\vec q} \cdot [\delta^{b \mu'}     \, {\vec k} \, c_{k}^{b \lambda'} +
                   \delta^{b \lambda'} \, {\vec p} \, c_{p}^{b \mu'}      ] \} \, 
/ \, q^{4},
\label{eq:MCT-v-c}
\eea
with ${\vec p} = {\vec q} - {\vec k}$.
Here, the direct correlation function is defined via the 
Ornstein-Zernike equation for a mixture~\cite{Hansen86},
$\rho c_{q}^{ab} = \delta^{ab} - [S_{q}^{-1}]^{ab}$.
Now, let us turn on the intramolecular constraints between
constituent atoms.
This amounts to replacing the direct correlation function
$c_{q}^{ab}$ for a mixture of spherical particles with the
one for molecular systems defined via the site-site
Ornstein-Zernike equation~\cite{Chandler72,Hansen86},
$\rho c_{q}^{ab} = [w_{q}^{-1}]^{ab} - [S_{q}^{-1}]^{ab}$.
Here enter the intramolecular correlation functions $w_{q}^{ab}$
describing the constraints. 
The so obtained equations for a mixture contain a frequency
matrix ${\bf \Omega}_{q}^{2}$ which reflects the $3n$ independent
degrees of freedom of the molecule.
In particular, ${\bf \Omega}_{q}^{2}$ gets an $n$-fold degenerate
eigenvalue zero for $q = 0$ due to the $n$ particle-number-conservation
laws for the $n$ species.
The used classical theory cannot account for the fact that
vibrational degrees of freedom are frozen out at sufficiently
low temperature because of quantum effects.
To repair this shortcoming, we make the assumption
that in the regime of interest the rigidity of the
molecule can be accounted for by replacing the classical
flexible-molecule value for ${\bf \Omega}_{q}^{2}$ 
in front of the convolution integral in Eq.~(\ref{eq:GLE-v-a})
by the formula in Eq.~(\ref{eq:GLE-v-b}).
The matrix ${\bf \Omega}_{q}^{2}$ for a rigid molecule
exhibits only one eigenvalue zero for $q = 0$,
since there is only one independent conservation law for the 
number of molecules.

The MCT equations for the tagged-molecule density correlator
${\bf F}_{q,s}(t)$ defined in Eq.~(\ref{eq:Fs-def}) can be obtained
similarly, and only the resulting equations shall be quoted. 
The exact Zwanzig-Mori equation reads
\bea
& &
\partial_{t}^{2} {\bf F}_{q,s}(t) + 
{\bf \Omega}_{q,s}^{2} \, {\bf F}_{q,s}(t) 
\nonumber \\
& &
\quad
+ 
{\bf \Omega}_{q,s}^{2} 
\int_{0}^{t} dt' \, {\bf m}_{q,s}(t-t') \, \partial_{t'} {\bf F}_{q,s}(t') = 
{\bf 0},
\label{eq:GLE-u}
\eea
where the characteristic frequency matrix is given as in 
Eq.~(\ref{eq:GLE-v-b}) by
${\bf \Omega}_{q,s}^{2} = q^{2} \, {\bf J}_{q} \, {\bf w}^{-1}_{q}$.
The expression for the relaxation kernel can be formulated 
as mode-coupling functional
$\mbox{\boldmath ${\cal F}$}_{q,s}$:
\begin{equation}
m^{ab}_{q,s}(t) = {\cal F}^{ab}_{q,s}[{\bf F}_{s}(t), {\bf F}(t)]. 
\label{eq:MCT-u-a}
\end{equation}
The explicit expression for the functional $\mbox{\boldmath ${\cal F}$}_{q,s}$
reads, with ${\vec p} = {\vec q} - {\vec k}$, 
\bea
& &
{\cal F}^{ab}_{q,s}[\tilde{\textit{\textbf f}}_{s}, \tilde{\textit{\textbf f}} \,] =
\frac{\rho}{(2 \pi)^{3}}
\sum_{c} \frac{w^{ac}_{q}}{q^{2}}
\sum_{\lambda, \mu} \int d{\vec k} \,
\biggl( \frac{ {\vec q} \cdot {\vec p} }{q} \biggr)^{2} \,
c_{p}^{c \lambda} \, c_{p}^{b \mu} 
\nonumber \\
& &
\qquad \qquad \qquad \qquad \qquad \qquad \qquad \quad
\times \, 
\tilde{f}^{cb}_{k,s} \, \tilde{f}^{\lambda \mu}_{p}.
\label{eq:MCT-u-b}
\eea

Let us note some mathematical results valid for the MCT formulated
above.
First of all, there is a solution ${\bf F}_{q}(t)$ 
of the nonlinear equations of motion for all times $t$.
This solution is uniquely fixed by the initial conditions
${\bf F}_{q}(t = 0) = {\bf S}_{q}$ and
$\partial_{t} {\bf F}_{q}(t = 0) = {\bf 0}$.
For every finite time interval, the solution depends smoothly
on the numbers $\Omega_{q}^{2 \, ab}$ and
$V_{\lambda \lambda' \mu \mu'}^{ab}({\vec q}; {\vec k}, {\vec p} \,)$.
The solutions are correlation functions in the sense that they can
be Laplace transformed to functions having a spectral representation
with a spectrum ${\bf F}''_{q}(\omega)$ which is a positive definite
matrix.
The matrix of long-time limits 
${\bf F}_{q} = \lim_{t \to \infty} {\bf F}_{q}(t)$ obeys the set of
implicit equations defined by the mode-coupling functional
$\mbox{\boldmath ${\cal F}$}_{q}$:
\begin{equation}
{\bf F}_{q} \, [{\bf S}_{q} - {\bf F}_{q}]^{-1} = 
\mbox{\boldmath ${\cal F}$}_{q}[{\bf F}]. 
\label{eq:DW-v}
\end{equation}
Let us remember that there is a semi-ordering in the space of real
symmetric $n$-by-$n$ matrices, 
${\bf A} > {\bf B}$, defined by
${\bf A} - {\bf B}$ to be positive definite. 
With this notation, the maximum theorem holds:
if $\hat{\bf F}_{q}$ is a solution of Eq.~(\ref{eq:DW-v}),
i.e. if 
$\hat{\bf F}_{q} [{\bf S}_{q} - \hat{\bf F}_{q}]^{-1} = 
\mbox{\boldmath ${\cal F}$}_{q}[\hat{\bf F}]$,
then 
$\hat{\bf F}_{q} \le {\bf F}_{q}$.
If an iteration sequence ${\bf F}_{q}^{(j)}$, $j = 0, 1, \cdots$,
is defined by
${\bf F}_{q}^{(j+1)} [{\bf S}_{q} - {\bf F}_{q}^{(j+1)}]^{-1} = 
\mbox{\boldmath ${\cal F}$}_{q}[{\bf F}^{(j)}]$
starting from ${\bf F}_{q}^{(0)} = {\bf S}_{q}$, then
${\bf F}_{q}^{(j+1)} < {\bf F}_{q}^{(j)}$ and
$\lim_{j \to \infty} {\bf F}_{q}^{(j)} = {\bf F}_{q}$.
If the Jacobian of Eq.~(\ref{eq:DW-v}) does not have a vanishing
eigenvalue, the long-time limit ${\bf F}_{q}$ depends smoothly
on the coupling coefficients
$V_{\lambda \lambda' \mu \mu'}^{ab}({\vec q}; {\vec k}, {\vec p} \,)$.
A singularity can occur only if an eigenvalue vanishes.
The subtlest property is that such vanishing 
eigenvalue is non-degenerate.
Hence, using the terminology of Arnold~\cite{Arnold86},
all possible singularities are bifurcations of the cuspoid type
$A_{\ell}$, $\ell = 2, 3, \cdots$.
The generic singularity for changes of a single control
parameter is, as for the MCT of simple systems,
a fold bifurcation $A_{2}$.
It is then obvious, that all universal results for
simple systems are valid also 
for the MCT for molecular systems formulated above.
For example, Eq.~(\ref{eq:fq-N-near-phic}) holds
with $f_{q}^{Nc}$ and $h_{q}^{N}$ replaced by positive definite 
matrices.
The proofs of the cited results of this paragraph shall not be 
described here for brevity, 
since essentially the same issues for matrices of 
density correlators 
have been independently discussed and proved by 
Franosch and Voigtmann~\cite{Thomas-stability}.

\end{document}